\documentclass[twocolumn]{aastex7}
\usepackage{ulem}

\newcommand\subs[1]{\textsubscript{#1}}
\newcommand\sups[1]{\textsuperscript{#1}}
\newcommand\rh[1]{\textcolor{black}{{\textit{r}\subs{\textit{H}}}#1}}

\newcommand\Ju[1]{\textcolor{black}{{\textit{J}}#1}}

\newcommand\kms[1]{\textcolor{black}{{km\,s$^{-1}$}#1}}
\newcommand\ps[1]{\textcolor{black}{{s$^{-1}$}#1}}
\newcommand\brh[1]{\textcolor{black}{{$\beta_p$(\rh{}=1 au)}#1}}

\definecolor{gold}{rgb}{0.64,0.54,0.29}

\received{2025 November 7}
\revised{2026 February 24}
\accepted{2026 March 2}
\submitjournal{The Planetary Science Journal}

\shorttitle{ALMA Imaging of Comet C/2017 K2 (PanSTARRS)}
\shortauthors{Roth et al.}

\begin{document}


\title{The Evolution in Coma Molecular Composition of Comet C/2017 K2 (PanSTARRS) Across the H$_2$O Sublimation Zone: ALMA Imaging of an H$_2$O-Dominated Coma}

\correspondingauthor{Nathan X. Roth}
\email{nathaniel.x.roth@nasa.gov}

\author[0000-0002-6006-9574]{Nathan X. Roth}
\affiliation{Solar System Exploration Division, Astrochemistry Laboratory Code 691, NASA Goddard Space Flight Center, 8800 Greenbelt Rd, Greenbelt, MD 20771, USA}
\affiliation{Department of Physics, American University, 4400 Massachusetts Ave NW, Washington, DC 20016, USA}
\email{nathaniel.x.roth@nasa.gov}

\author[0000-0001-7694-4129]{Stefanie N. Milam}
\affiliation{Solar System Exploration Division, Astrochemistry Laboratory Code 691, NASA Goddard Space Flight Center, 8800 Greenbelt Rd, Greenbelt, MD 20771, USA}
\email{stefanie.n.milam@nasa.gov}

\author[0000-0001-8233-2436]{Martin A. Cordiner}
\affiliation{Solar System Exploration Division, Astrochemistry Laboratory Code 691, NASA Goddard Space Flight Center, 8800 Greenbelt Rd, Greenbelt, MD 20771, USA}
\affiliation{Department of Physics, The Catholic University of America, 620 Michigan Ave., N.E. Washington, DC 20064, USA}
\email{martin.cordiner@nasa.gov}

\author[0000-0001-9479-9287]{Anthony J. Remijan}
\affiliation{National Radio Astronomy Observatory, 520 Edgemont Rd, Charlottesville, VA 22903, USA}
\email{aremijan@nrao.edu}

\author{Dominique Bockelée-Morvan}
\affiliation{LIRA, Observatoire de Paris, Université PSL, CNRS, Sorbonne Université, Université Paris Cité, 5 place Jules Janssen, 92195 Meudon, France}
\email{dominique.bockelee@obspm.fr}

\author[0000-0003-2414-5370]{Nicolas Biver}
\affiliation{LIRA, Observatoire de Paris, Université PSL, CNRS, Sorbonne Université, Université Paris Cité, 5 place Jules Janssen, 92195 Meudon, France}
\email{nicolas.biver@obspm.fr}

\author[0000-0002-1545-2136]{Jérémie Boissier}
\affiliation{Institut de Radioastronomie Millimetrique, 300 rue de la Piscine, F-38406
Saint Martin d'Heres, France}
\email{boissier@iram.fr}

\author[0000-0001-6752-5109]{Steven B. Charnley}
\affiliation{Solar System Exploration Division, Astrochemistry Laboratory Code 691, NASA Goddard Space Flight Center, 8800 Greenbelt Rd, Greenbelt, MD 20771, USA}
\email{steven.b.charnley@nasa.gov}

\author[0000-0001-6567-627X]{Charles E.\ Woodward}
\affiliation{Minnesota Institute for Astrophysics, School of
Physics and Astronomy, 116 Church Street, S.E., University of
Minnesota, Minneapolis, MN 55455, USA}
\email{chickw024@gmail.com}

\author[0009-0009-9925-7001]{Lillian X. Hart}
\affiliation{Department of Physics, University of Maryland, College Park, MD}
\email{lillianxh1@gmail.com}

\author[0009-0005-1065-4109]{Timothy N. Proudkii}
\affiliation{Division of Geological and Planetary Sciences, California Institute of Technology, Pasadena, CA 91125, USA}
\affiliation{Department of Physics, Virginia Tech, 850 West Campus Drive, Blacksburg, VA 24061, USA}
\email{timproudkii@caltech.edu}




\begin{abstract}

We report a survey of molecular emission from cometary volatiles using the Atacama Large Millimeter/Submillimeter Array (ALMA) toward comet C/2017 K2 (PanSTARRS) carried out on UT 2022 September 21, 22, and 23 at a heliocentric distance (\rh{}) of 2.1 au. These measurements of HCN, CS, CO, CH$_3$OH, and H$_2$CO (along with continuum emission from dust) sampled molecular chemistry in C/2017 K2 at the inner edge of the H$_2$O sublimation zone, the region from \rh{} = 2-3 au where H$_2$O begins vigorously subliming and increasingly dominating comet activity, discerning parent from daughter or extended source species. This work presents spectrally integrated flux maps, production rates, and parent scale lengths for each molecule. CH$_3$OH, CO, and HCN were produced within $\sim$250 km of the nucleus, potentially including contributions from sublimation of icy grains. CS was consistent with production from CS$_2$ photolysis, and H$_2$CO required production from extended sources in the coma. An ortho-to-para ratio OPR=$2.9\pm0.4$ for H$_2$CO was derived from simultaneously measured transitions of each spin species. The continuum was extended and spatially resolved, consistent with thermal emission from dust in the coma. Analysis of the continuum visibilities provided an upper limit on the nucleus diameter $d<6.6$ km and coma dust masses of $1.2-2.4\times10^{11}$ kg.

\end{abstract}



\section{Introduction} \label{sec:intro}
Comets provide a window into the early solar system. They formed in the cold disk midplane of the protosolar disk during the era of planet formation and were subsequently scattered to the Kuiper disk or the Oort cloud. Cryogenically preserved for the last $\sim$4.5 Gyr in the  cold outer solar system, they may serve as ``fossils'' of solar system formation, with the volatile composition of their nuclei reflecting the chemistry and prevailing conditions present where and when they formed \citep{Bockelee2004,Mumma2011a,Bockelee2017}.  

As comets (particularly from the Oort cloud) pass into the inner solar system, they will cross into multiple sublimation zones. These sublimation zones correspond to the heliocentric distances (\rh{}) at which sufficient solar insolation becomes available to activate sublimation of nucleus ices with differing levels of volatility. Consequently, different volatiles drive cometary activity over the course of an orbit depending on the \rh{} at a given time. Previous studies of comets at large \rh{} have suggested that CO sublimation can begin at up to tens of au \citep{Meech2009}, followed by CO$_2$ near 6--8 au and H$_2$O around 2--3 au \citep{Kelley2016}. Measurements of comet C/1995 O1 (Hale-Bopp) preformed with IRAM 30 m, the Plateau de Bure interferometer, and SEST 15 m detected CO at 6.8 au pre-perihelion and as far as 14 au post-perihelion \citep{Biver2002}. These monthly observations demonstrated that activity in Hale-Bopp was driven by CO for \rh{} $>$ 3 au, followed by H$_2$O-driven activity. Similarly, CH$_3$OH and HCN (with volatility lower than CO but higher than H$_2$O) were overabundant at large \rh{} before approaching average values closer to the Sun, whereas CS, H$_2$CO, and HNC underwent steep increases in production with \rh{}.

Comprehensive studies of the primary drivers of cometary activity (H$_2$O, CO, and CO$_2$) providing production rates over a wide range of \rh{} are few and far between. CO$_2$ is completely unobservable from the ground, leaving its contributions to cometary activity and nucleus ice content little understood. What little is known about cometary CO$_2$ comes from rendezvous missions at relatively low \rh{} ($<$4 au) \citep{AHearn2011,Hassig2015}, spacecraft remote sensing surveys \citep{Ootsubo2012,Reach2013}, and pointed targetings \citep{Woodward2025,Snodgrass2025,Cordiner2025}. The few studies targeting all three molecules have shown that their abundances vary widely among comets \citep{Ahearn2012,Feaga2014,Harrington2022}. 

In particular, the JWST results have demonstrated the importance of CO$_2$ in driving coma activity beyond the H$_2$O sublimation zone, with CO$_2$-dominated comae found for comet C/2024 E1 (Wierzchos) at \rh{} = 7 au and for interstellar object 3I/ATLAS at 3.3 au \citep{Snodgrass2025,Cordiner2025}. On the other hand, JWST observations of comet C/2017 K2 (PanSTARRS; the target of this study) at \rh{} = 2.35 au found an H$_2$O-dominated coma with CO$_2$ and CO on similar footing as trace species \citep[relative abundances of CO$_2$/H$_2$O = 14\% and CO/H$_2$O = 8\%, respectively;][]{Woodward2025}, highlighting how overall coma chemistry changes with \rh{}.

Here we report ALMA Cycle 8 observations of comet C/2017 K2 (PanSTARRS; hereafter K2). K2 was already active upon its discovery at 16 au, with activity consistent with CO sublimation \citep{Meech2017a}. Analysis of pre-discovery observations revealed activity at \rh{} = 23.7 au, and Monte-Carlo analysis of photometry suggested that activity began as distantly as \rh{} = 35 au, making C/2017 K2 a ultra-distantly active comet \citep{Jewitt2017,Jewitt2021}. Observations of comets at such large \rh{} are exceedingly rare, where it is far too cold for H$_2$O sublimation or phase transitions to produce a coma, and activity is likely driven by sublimation of supervolatile ices. Follow-up observations confirmed CO-dominated activity at \rh{} = 6.7 au \citep{Yang2021}. With its large discovery distance and high intrinsic brightness, studies of K2 afforded an opportunity to reveal how the abundances, outgassing mechanisms, and spatial distributions of coma molecules evolved as it crossed sublimation zones pre-perihelion.

Such studies of ultra-distantly active comets with perihelia in the inner solar system (\rh{} $<\sim3$ au) afford an opportunity to better understand those that remain at greater distances from the Sun and provide a bridge to place studies of distant comets into context with the larger sample of comets measured closer to the Sun. Observations with NASA IRTF, Keck II, and JWST cataloged its dust and volatile composition measured at near- and mid-IR wavelengths. IRTF and Keck provided serial pre-perihelion observations between \rh{} = 3.15 and 2.35 au, first detecting H$_2$O at \rh{} = 2.5 au and finding generally super-enriched abundances of trace volatiles \citep{Ejeta2025}. JWST observations at \rh{} = 2.35 au found evidence for H$_2$O sublimation from icy grains in the coma, added $^{13}$CO$_2$, $^{12}$CO$_2$, and OCS to the detected volatile inventory, and found strong evidence for the presence of polycyclic aromatic hydrocarbons (PAHs) in the coma \citep{Woodward2025}. 

Our ALMA observations monitored K2 as it approached and crossed into the H$_2$O sublimation zone from \rh{} = 4.1 au -- 2.1 au, with each epoch scheduled as \rh{} decreased by 0.5 au. Results are reported here from the final epoch at \rh{} = 2.1 au in an H$_2$O-dominated coma and serve as a compositional baseline for comparison against measurements at larger 
\rh{} when H$_2$O was not yet vigorously subliming. Subsequent publications will report our remaining epochs at \rh{} = 4.1 au, 3.5 au, 3.0 au, and 2.5 au, cataloging how K2's molecular composition evolved from beyond to within the H$_2$O sublimation zone, transitioning from CO-dominated to H$_2$O-dominated outgassing. 

Secure detections of molecular emission were identified from HCN, CS, CO, CH$_3$OH, and H$_2$CO, as well as continuum emission from dust and the nucleus. We report production rates, mixing ratios (relative to H$_2$O and CO), and spatial maps of the detected species. Section~\ref{sec:obs} discusses the observations and data reduction. Section~\ref{sec:results} presents our results from these data. Section~\ref{sec:modeling} details our modeling approach to analyze spectral line and continuum emission, and Section~\ref{sec:discussion} places our results into context with studies of C/2017 K2 at other wavelengths and the larger comet population.
\section{Observations and Data Reduction} \label{sec:obs}
Comet K2 is an Oort cloud comet and one of only four comets discovered with activity at \rh{} $>$ 20 au \citep{Meech2017a}. There is debate regarding whether K2 is dynamically new \citep[entering the inner solar system for the first time;][]{Krolikowska2018}. During its 2022 apparition, K2 reached perihelion (\textit{q} = 1.79 au) on UT 2022 December 19 and passed closest to the Earth ($\Delta$\subs{min} = 1.81 au) on UT 2022 July 14. We conducted pre-perihelion observations toward K2 on UT 2022 September 21-23 during Cycle 8 using the ALMA 12 m array with the Band 7 receiver, covering frequencies between 290.15 and 364.16 GHz ($\lambda$ = 0.82 -- 1.03 mm) in seventeen non-contiguous spectral windows ranging from 234--469 MHz wide (for the 357 GHz setting) to 117--2000 MHz wide (for the 297 GHz setting) to 117--468 MHz wide (for the 349 GHz setting). The observing log is shown in Table~\ref{tab:obslog}. We tracked the comet position using JPL Horizons ephemerides (JPL \#82). Three correlator settings were employed to simultaneously sample spectral lines from multiple molecules and continuum. Mean precipitable water vapor at zenith (zenith PWV) ranged from 0.73--0.84 mm. Quasar observations were used for bandpass and phase calibration, as well as calibrating K2's flux scale. The spatial scale (the range in semi-minor and semi-major axes of the synthesized beam) was 0$\farcs$49 -- 0$\farcs$94 and the channel spacing was 15.6 MHz for continuum windows, 122 kHz for the HCN spectral line window, and 244 kHz for all other spectral line windows. An averaging width of two spectral channels was employed at the correlator, resulting in a spectral resolution of 0.12 \kms{} for HCN and 0.25 \kms{} for all other species. 

The data were flagged, calibrated, and imaged using standard routines in Common Astronomy Software Applications (CASA) package version 6.4.1 \citep{CASA2022}. We used the \texttt{TCLEAN} task in CASA for image reconstruction. We deconvolved the point-spread function with the Högbom algorithm, using natural visibility weighting, a 9$\arcsec$ diameter mask centered on the peak continuum position (spanning projected radial distances of 7700 km), and a flux threshold of twice the rms noise. The deconvolved images were then convolved with the synthesized beam and corrected for the (Gaussian) response of the ALMA primary beam. We transformed the images from astrometric coordinates to projected cometocentric distances, with the location of the peak continuum flux chosen as the origin, which was in good agreement with the comet's predicted ephemeris position ($<0\farcs450$ offset between the peak continuum position and the phasecenter). 


\begin{deluxetable*}{ccccccccccccc}
\tablenum{1}
\tablecaption{Observing Log\label{tab:obslog}}
\tablewidth{0pt}
\tablehead{
\colhead{Date} & \colhead{UT Time} & \colhead{\textit{T}\subs{int}} &
\colhead{\textit{r}\subs{H}} & \colhead{$\Delta$} & \colhead{$\phi_\mathrm{STO}$} & \colhead{$\psi_{\sun}$} & \colhead{$\nu$} & 
\colhead{\textit{N}\subs{ants}}  & \colhead{Baselines} & \colhead{PWV} & \colhead{$\theta$\subs{min}} & \colhead{$\theta$\subs{min}} \\
\colhead{(2022)} & \colhead{} & \colhead{(min)} & \colhead{(au)} & 
\colhead{(au)} & \colhead{($\degr$)} & \colhead{($\degr$)} & \colhead{(GHz)}  & \colhead{} & \colhead{(m)}  & \colhead{(mm)} & 
\colhead{($\arcsec$)} & \colhead{(km)}
}
\startdata
21--22 Sept & 23:27--00:07 & 33 & 2.12 & 2.34 & 25.3 & 106.5 & 349 & 43 & 15--500 & 0.73 &  0.88$\times$0.49 & 1493$\times$831 \\ 
22 Sept & 21:07--21:49 & 33 & 2.11 & 2.35 & 25.2 & 106.8 & 349 & 43 & 15--500 & 0.84 & 0.65$\times$0.49 & 1107$\times$835 \\
22 Sept & 22:08--22:44 & 28 & 2.11 & 2.35 & 25.2 & 106.8 & 297 & 43 & 15--500 & 0.81 & 0.83$\times$0.57 & 1414$\times$971 \\
22 Sept & 23:08--23:44 & 28 & 2.11 & 2.35 & 25.2 & 106.8 & 357 & 43 & 15--500 & 0.75 & 0.86$\times$0.49 & 1465$\times$835 \\
23--24 Sept & 23:28--00:04 & 28 & 2.10 & 2.36 & 25.1 & 107.1 & 357 & 43 & 15--500 & 0.73 & 0.94$\times$0.50 & 1608$\times$855 \\ 
\enddata
\tablecomments{\textit{T}\subs{int} is the total on-source integration time. \textit{r}\subs{H}, $\Delta$, and $\phi_\mathrm{STO}$, and $\psi_{\sun}$ are the heliocentric distance, geocentric distance,
phase angle (Sun--Comet--Earth), and position angle of the Sun--Comet radius vector, respectively, of K2 at the time of observations. $\nu$ is the mean frequency of each instrumental setting. \textit{N}\subs{ants}
is the number of antennas utilized during each observation, with the range of baseline lengths indicated for each. PWV is the mean precipitable water vapor at zenith during the observations. $\theta$\subs{min} is the angular resolution (synthesized beam) at $\nu$ given in arcseconds and in projected distance (km) at the geocentric distance of K2.} 
\end{deluxetable*}
\section{Results} \label{sec:results}
\begin{figure*}
\gridline{\fig{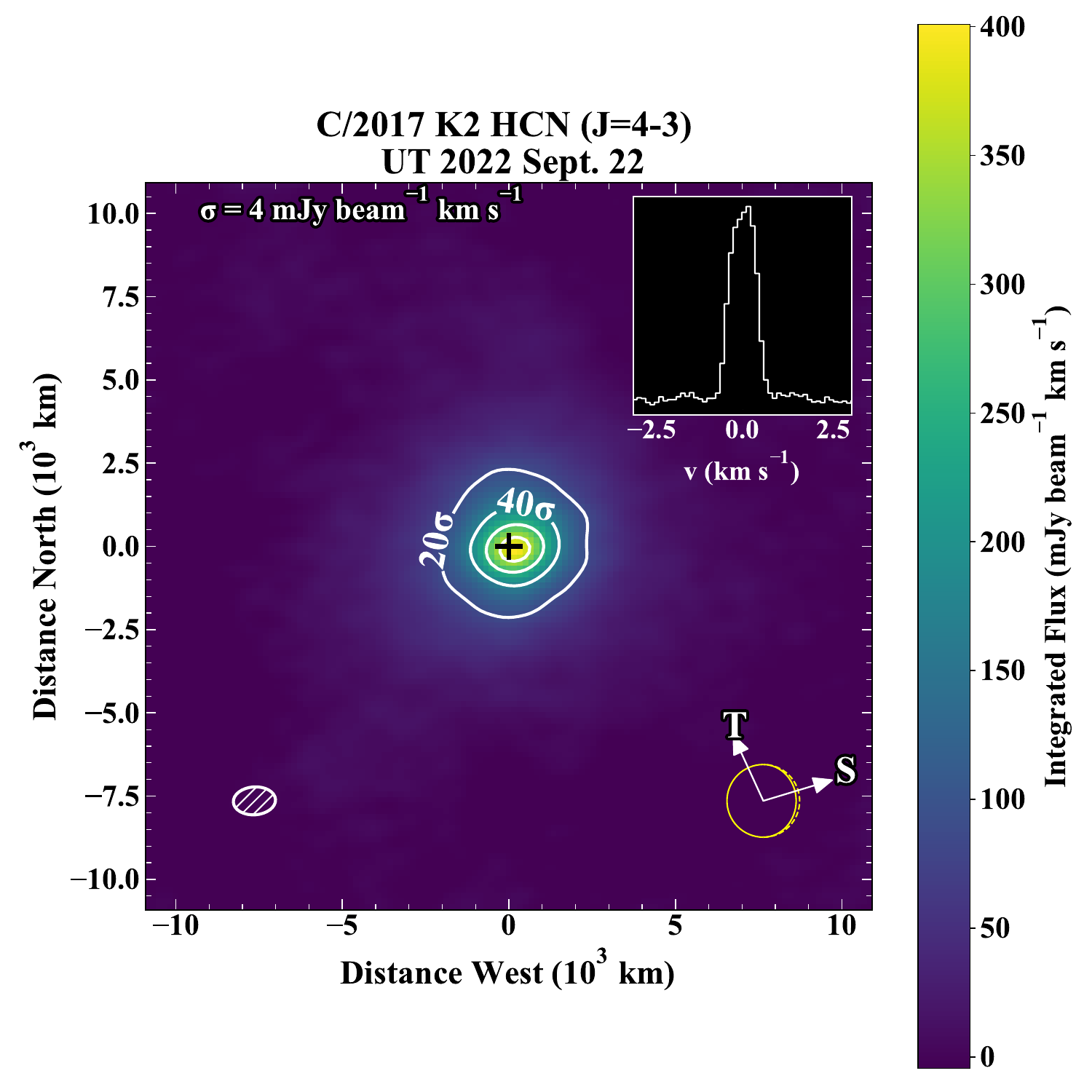}{0.45\textwidth}{(A)}
          \fig{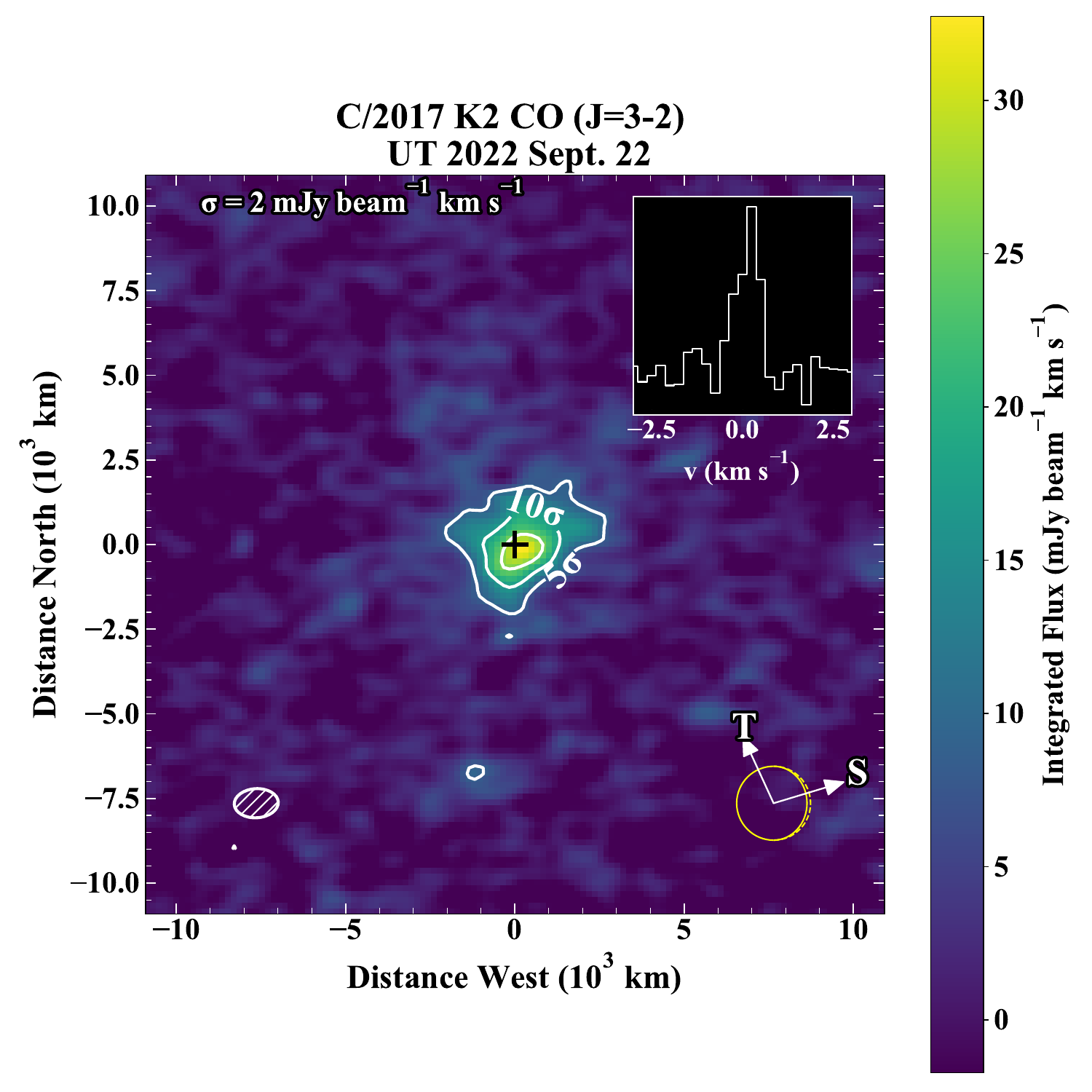}{0.45\textwidth}{(B)}
}
\gridline{\fig{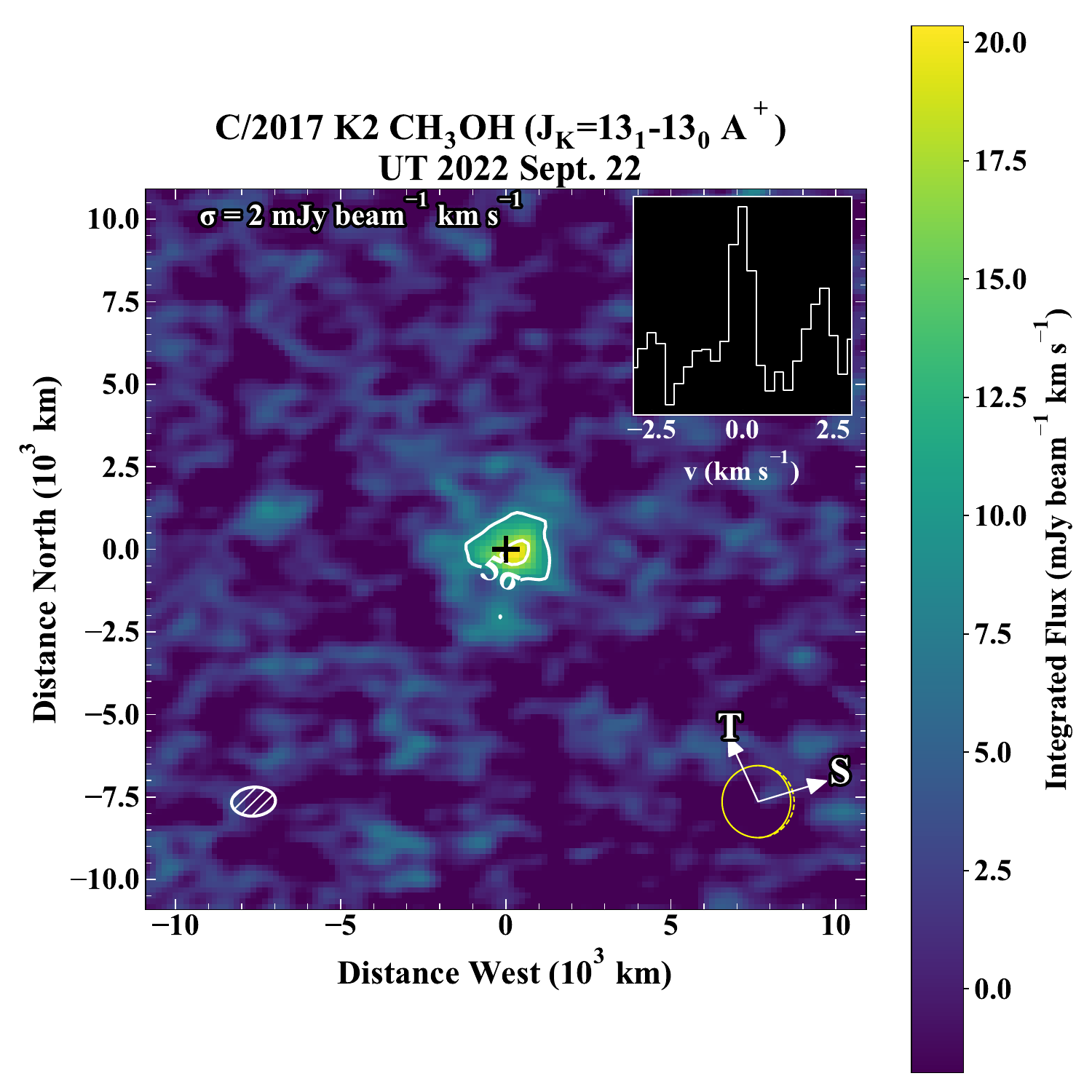}{0.45\textwidth}{(C)}
          \fig{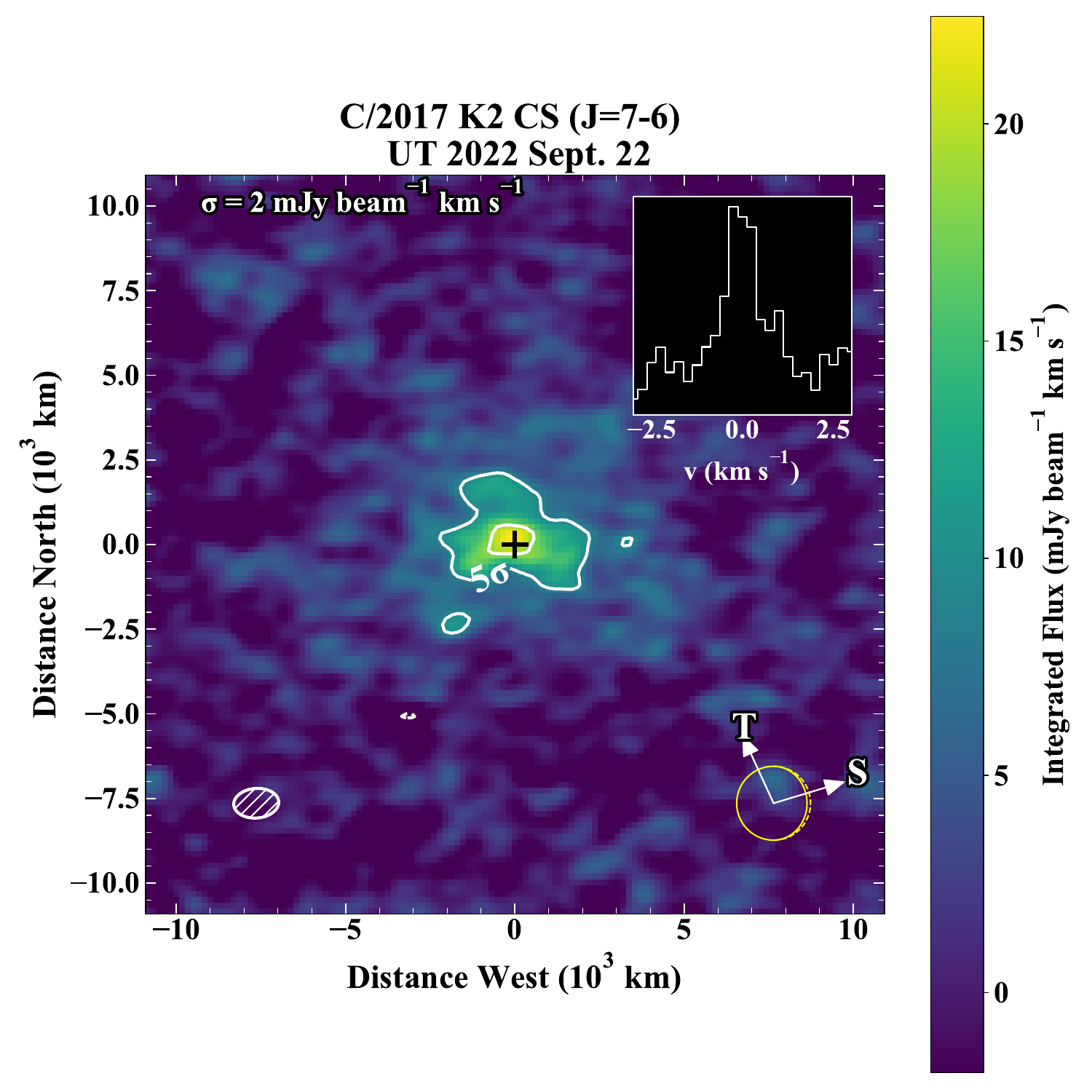}{0.45\textwidth}{(D)}
}
\caption{\textbf{(A)--(D).} Spectrally integrated flux maps for HCN, CO, CH$_3$OH, and CS in K2. Contour intervals in each map are given in multiples of the rms noise. The rms noise ($\sigma$, mJy beam$^{-1}$ km s$^{-1}$) is indicated in the upper left corner of each panel. Sizes and orientations of the synthesized beam (Table~\ref{tab:obslog}) are indicated in the lower left corner of each panel. The comet's observer-centered illumination ($\phi \sim$ 25$\degr$), as well as the direction of the Sun (S) and dust trail (T), are indicated in the lower right. The black cross indicates the position of the peak continuum flux. A spectrum extracted from a $10\arcsec$ diameter aperture centered on the nucleus (spanning projected radial nucleocentric distances of 8500 km) is shown in the upper right. Contours are in 20$\sigma$ increments for HCN and 5$\sigma$ increments for CO, CH$_3$OH, and CS.
\label{fig:hcn-maps}}
\end{figure*}

\begin{figure*}
\gridline{\fig{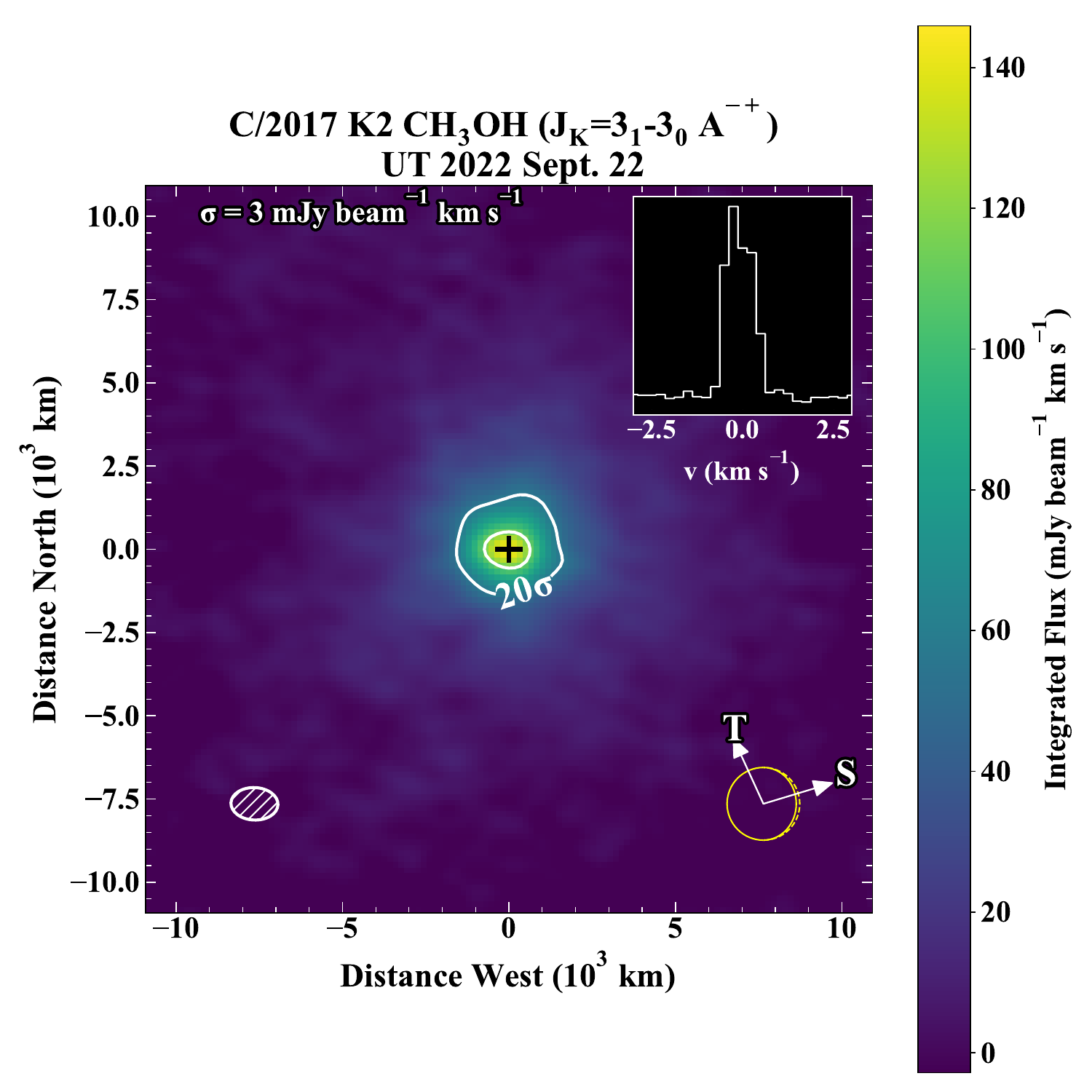}{0.45\textwidth}{(A)}
          \fig{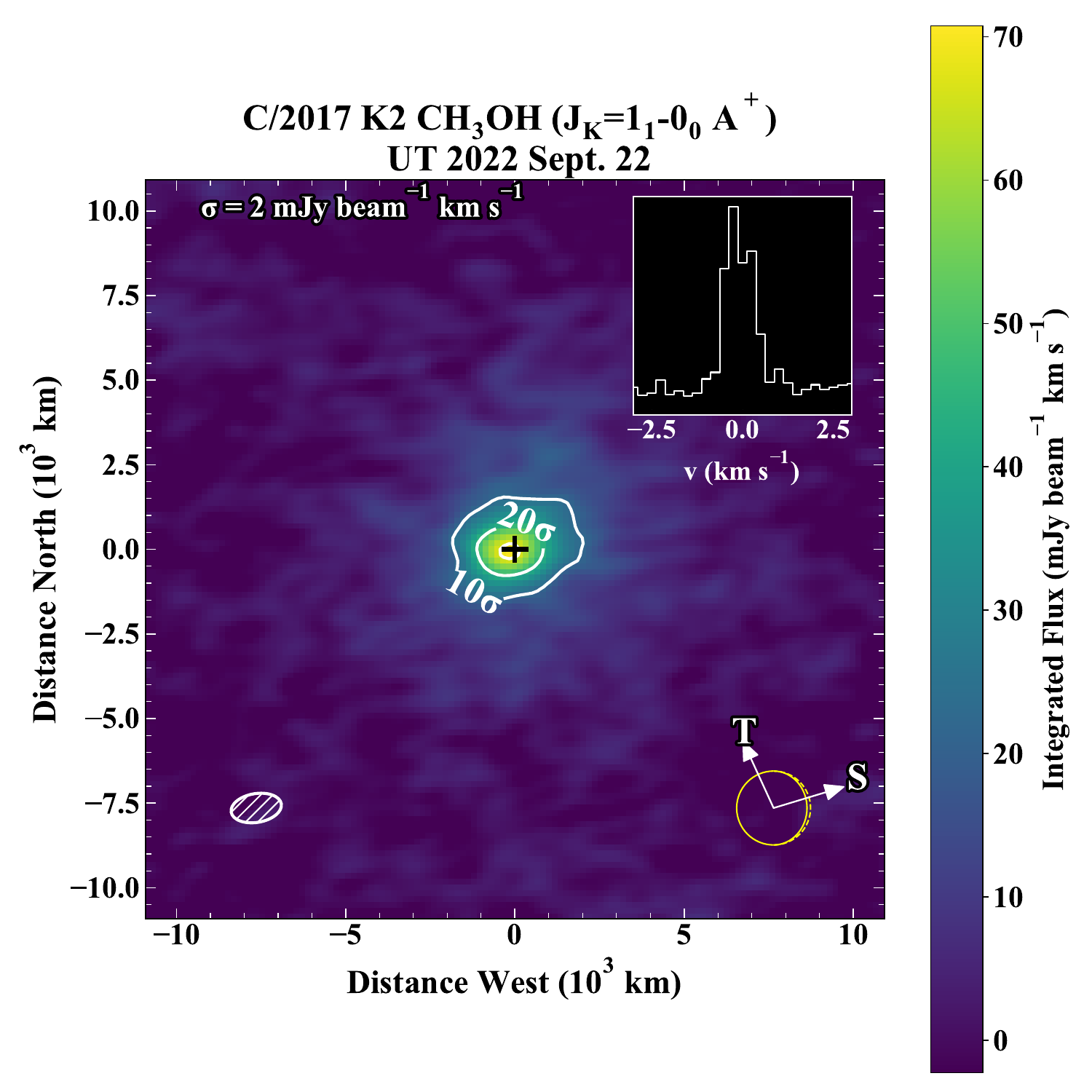}{0.45\textwidth}{(B)}
}
\gridline{\fig{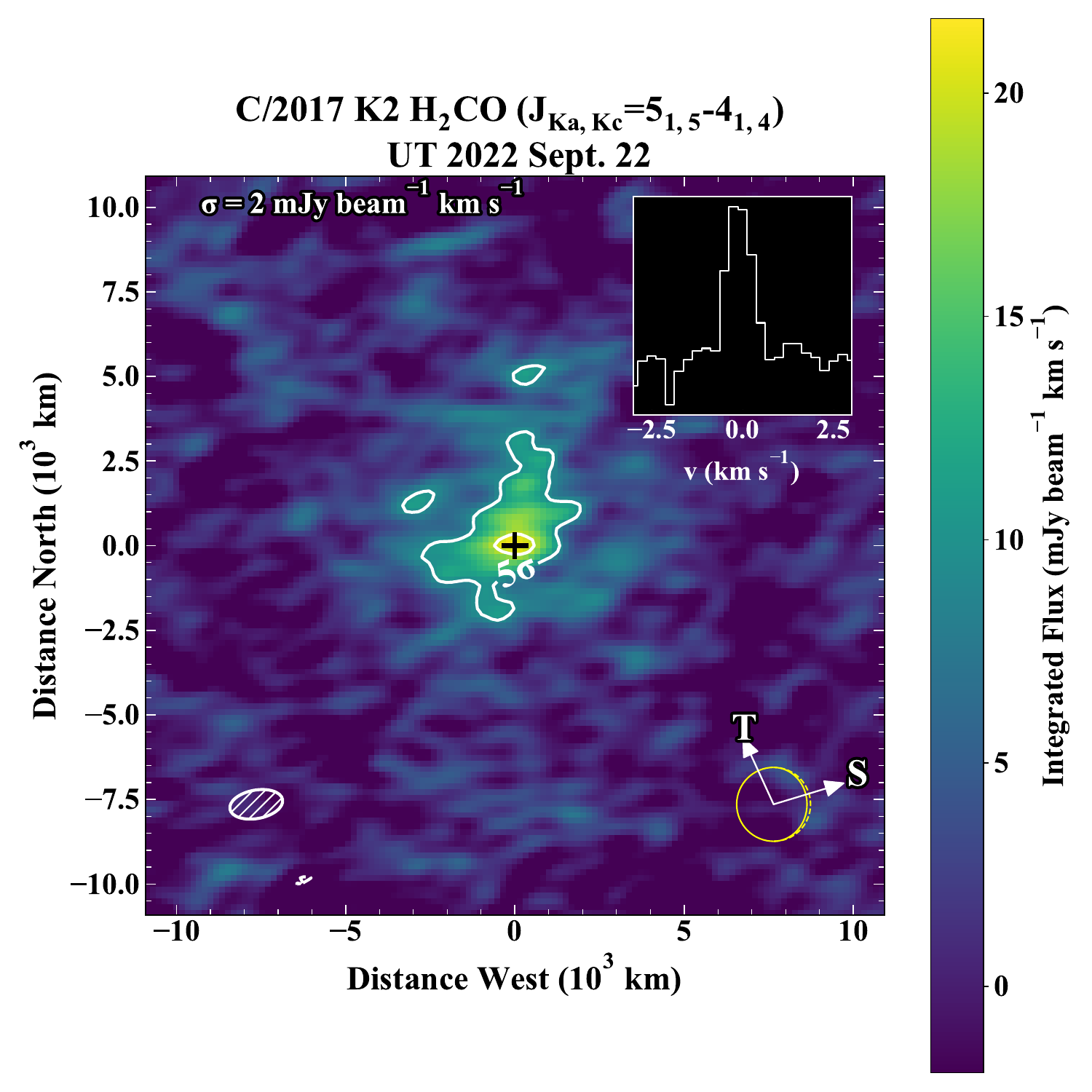}{0.45\textwidth}{(C)}
          \fig{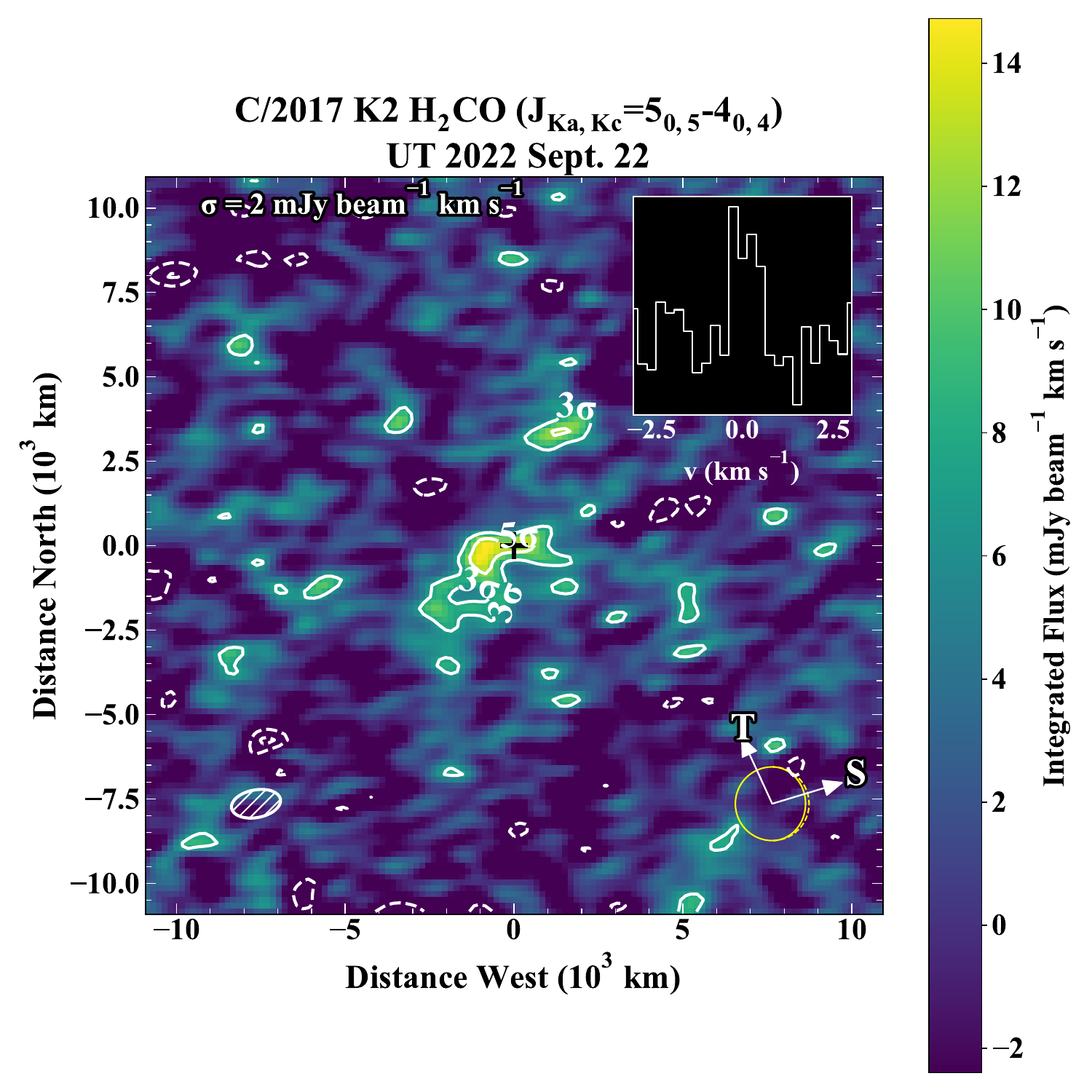}{0.45\textwidth}{(D)}
}
\caption{\textbf{(A)--(D).} Spectrally integrated flux map for CH$_3$OH and H$_2$CO in K2, with traces and labels as in Figure~\ref{fig:hcn-maps}. Contours are in 10$\sigma$ increments for CH$_3$OH and 5$\sigma$ increments for H$_2$CO ($J_{Ka,Kc}=5_{1,5}-4_{1,4}$). Contours are 3$\sigma$ and $5\sigma$ for H$_2$CO ($J_{Ka,Kc}=5_{0,5}-4_{0,4}$). In contrast to the other species, H$_2$CO showed asymmetric, clumpy, and  spatially extended emission.
\label{fig:tkin-maps}}
\end{figure*}

\begin{figure*}

\gridline{\fig{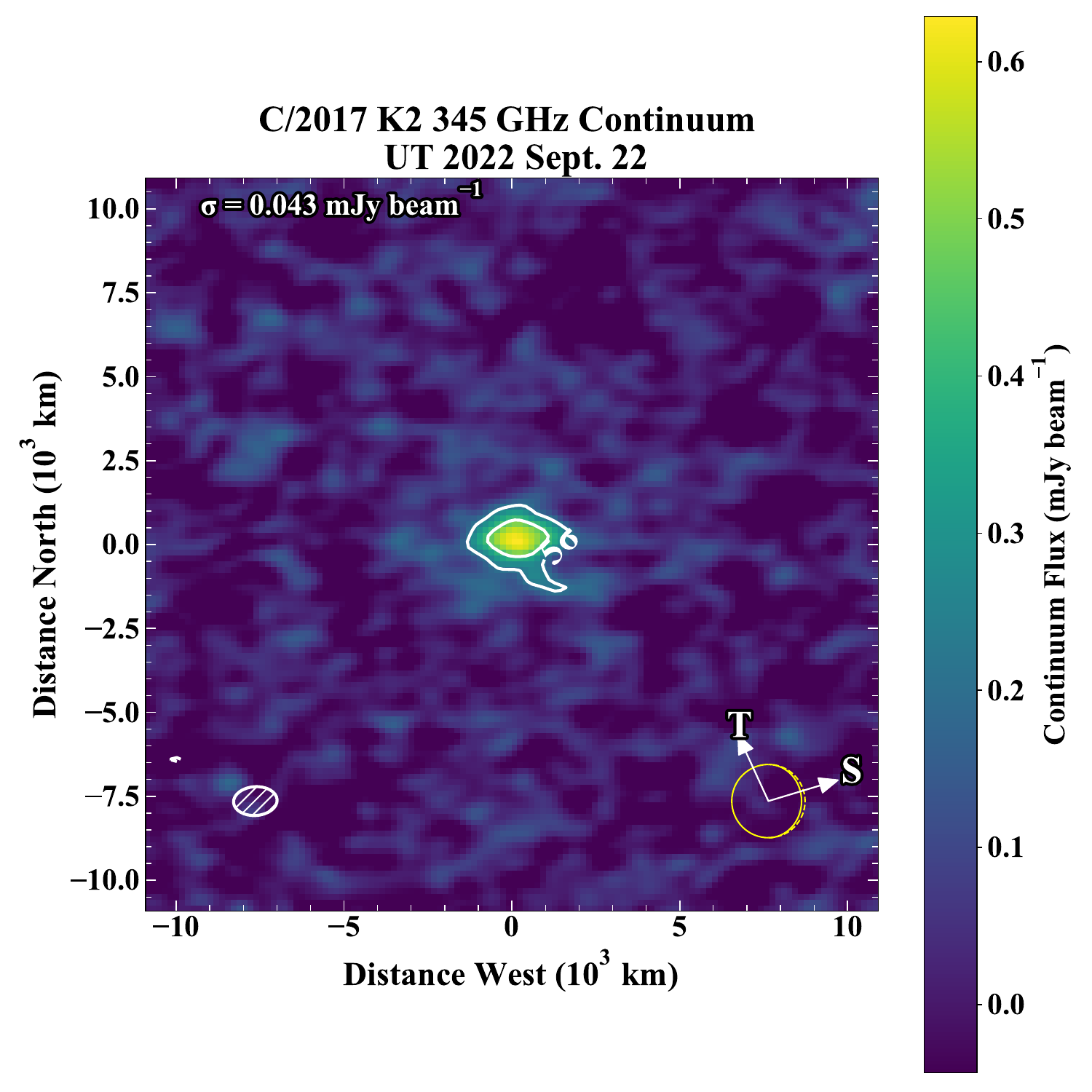}{0.45\textwidth}{(A)}
          \fig{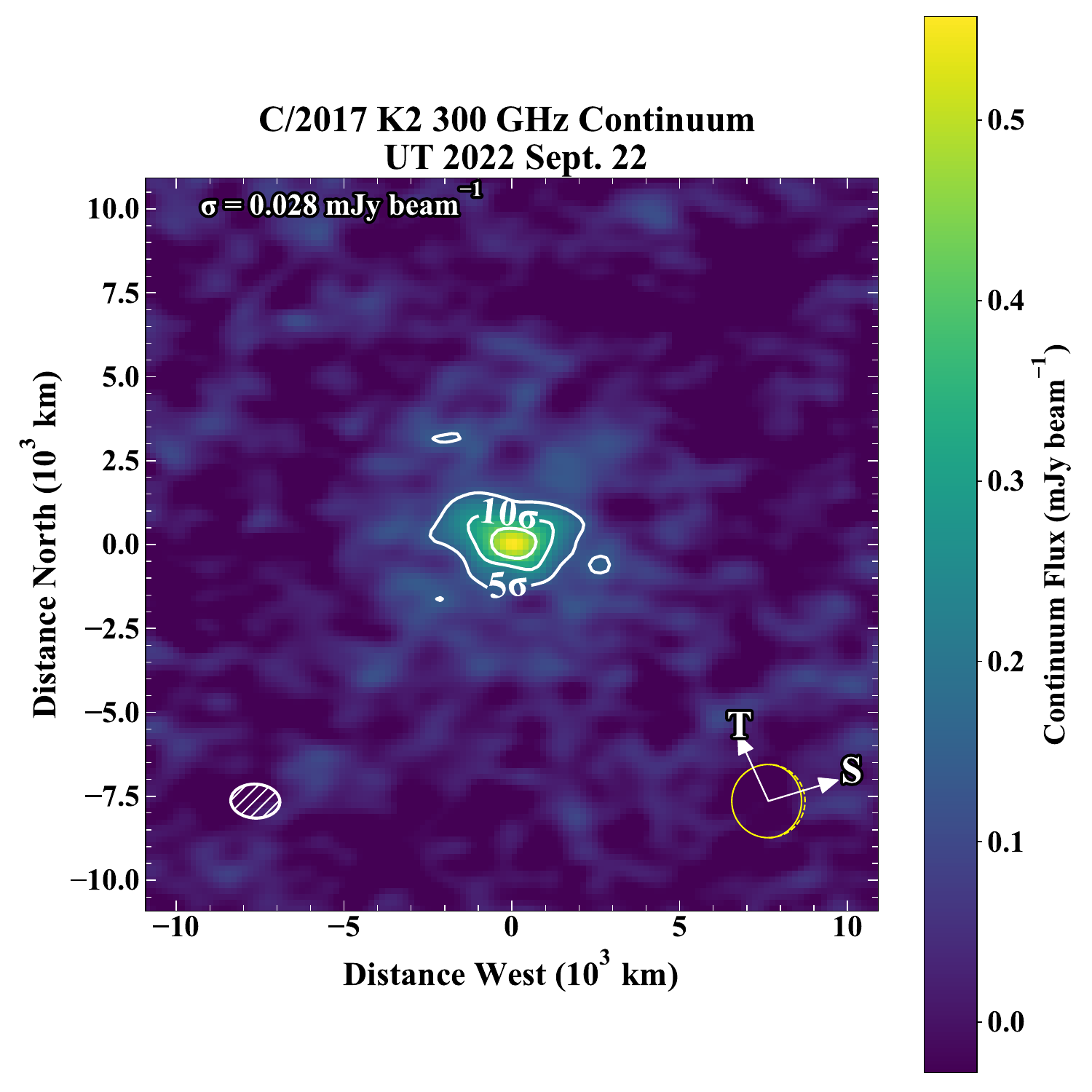}{0.45\textwidth}{(B)}
}
\gridline{\fig{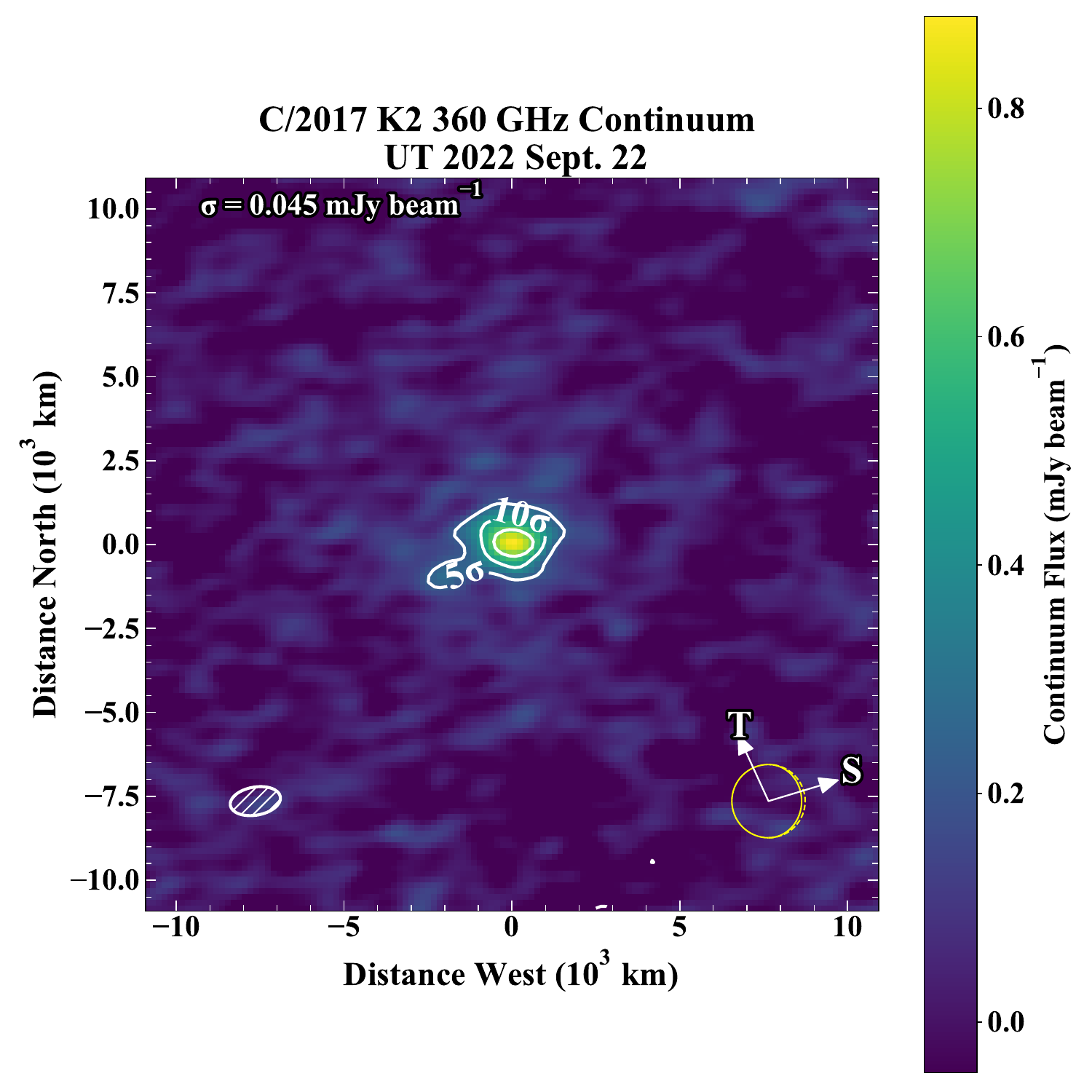}{0.45\textwidth}{(C)}
}
\caption{\textbf{(A)--(D).} Continuum flux maps in K2, with traces and labels as in Figure~\ref{fig:hcn-maps}. Contours are in 5$\sigma$ increments.
\label{fig:h2co-maps}}
\end{figure*}

HCN, CS, CO, and CH$_3$OH were detected on September 21; all of these plus H$_2$CO on September 22; and only CH$_3$OH and H$_2$CO on September 23. These detections included fourteen lines of CH$_3$OH spanning a wide range of excitation energies ($E_u$ = 16.8 K -- 227.5 K), thereby enabling a rigorous examination of the coma radial temperature profile, and five H$_2$CO lines, including simultaneous measures of one strong line each of ortho- and para-H$_2$CO. Continuum emission from dust/icy grains and the nucleus was detected on all dates. Extracted spectra and integrated flux maps are shown in Figures~\ref{fig:hcn-maps}, \ref{fig:tkin-maps}, and \ref{fig:h2co-maps}. Table~\ref{tab:lines} lists all detected molecular transitions.

The detected volatiles and continuum showed a range of spatial distributions. The continuum was compact about the nucleus compared to the molecular emission. HCN, CH$_3$OH, and (to a lesser extent) CO showed more symmetric distributions compared to CS and especially H$_2$CO, which were more asymmetric and clumpy. 

\begin{deluxetable*}{cccc}
\tablenum{2}
\tablecaption{Lines Detected in C/2017 K2 \label{tab:lines}}
\tablewidth{0pt}
\tablehead{
\colhead{Species} & \colhead{Transition$^{(a)}$} & \colhead{Frequency} & \colhead{\textit{E}\subs{u}}  \\
\colhead{} & \colhead{} & \colhead{(GHz)} & \colhead{(K)} 
}
\startdata
\hline
HCN & 4--3 & 354.505478 & 42.5 \\
CO & 3--2 & 345.795989 & 33.2 \\
CS & 7--6 & 342.882850 & 65.8 \\
$o$-H$_2$CO & $5_{1,5}$--$4_{1,4}$ & 351.768645 & 62.4 \\
$p$-H$_2$CO & $4_{0,4}$--$3_{0,3}$ & 290.623405 & 34.9 \\
            & $5_{0,5}$--$4_{0,4}$ & 362.736048 & 52.3 \\
            & $4_{2,3}$--$3_{2,2}$ & 291.237766 & 82.0 \\
            & $4_{2,2}$--$3_{2,1}$ & 291.948067 & 82.1 \\
CH$_3$OH & $1_1$--$0_0 A^+$ & 350.905100 & 16.8  \\
& $1_1$--$1_0 A^{-+}$ & 303.366921 & 16.9  \\
& $2_1$--$2_0 A^{-+}$ & 304.208348 & 21.6 \\
& $3_1$--$3_0 A^{-+}$ & 305.473491 & 28.6  \\
& $4_0$--$3_{-1} E$ & 350.687662 & 36.3  \\
& $6_0$--$5_0 A^+$ & 290.110637 & 48.7  \\
& $6_{-1}$--$5_{-1} E$ & 290.069747 & 54.3 \\
& $7_2$--$6_1 E$ & 363.739868 & 87.3  \\
& $6_4$--$5_4 A^-$ & 290.161348 & 129.1 \\
& $6_4$--$5_4 A^+$ & 290.161352 & 129.1 \\
& $6_{-4}$--$5_{-4} E$ & 290.162356 & 136.6 \\
& $6_{-5}$--$5_{-5} E$ & 290.117786 & 172.7 \\
& $13_0$--$12_1 A^+$ & 355.602945 & 211.0  \\
& $13_1$--$13_0 A^{-+}$ & 342.729796 & 227.5  \\
\enddata
\tablecomments{\sups{a} Quantum numbers are given as ($J'$--$J''$) for HCN, CO, and CS, as ($J'_{Ka'Kc'}$--$J''_{Ka''Kc''}$) for H$_2$CO, and as ($J'_{K'}$--$J''_{K''}$) for CH$_3$OH.
}
\end{deluxetable*}
\section{Modeling and Interpretation} \label{sec:modeling}
We chose to perform modeling of the interferometric spectra in the Fourier domain following \cite{Cordiner2023} to avoid the introduction of imaging artifacts arising from incomplete $uv$ sampling inherent to interferometric observations and the CLEAN algorithm itself. We first averaged the observed visibilities temporally. We then used the \texttt{vis\_sample} program \citep{Loomis2018} to take the Fourier transform of our radiative transfer model cubes using the same $uv$ coverage as our (temporally averaged) ALMA observations. We then performed least-squares fits of our radiative transfer models against the measured interferometric visibilities. We used the \texttt{lmfit} application of the Levenberg-Marquardt minimization technique and retrieved uncertainties on our optimized parameters from the diagonal elements of the covariance matrix, minimizing the residual for both the real and imaginary parts of the observed and modeled interferometric visibilities. 

\subsection{Analysis of Continuum Emission}\label{subsec:cont}
\begin{figure}
\plotone{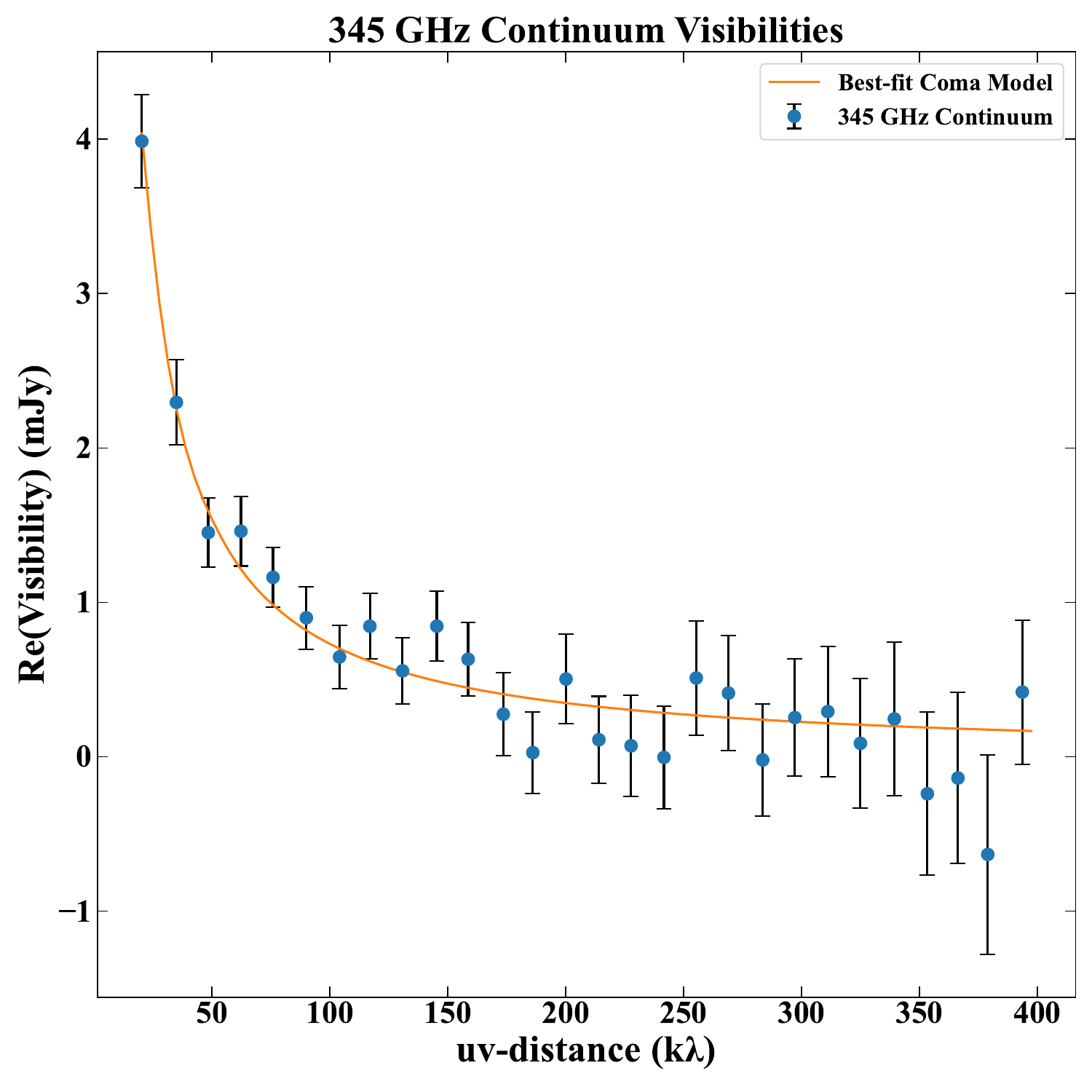}
\caption{345 GHz continuum visibilities in C/2017 K2, with best-fit coma model overlaid. The $uv$ distances have been plotted in units of k$\lambda$.
\label{fig:contVis}}
\end{figure}

We temporally averaged the ALMA continuum measurement set, then spatially averaged the continuum visibility amplitudes in 12 m $uv$ radius bins. We then measured the real part of the visibility amplitudes in the Fourier domain on the shortest baselines ($uv$-distances $\rho<20$ m) in each setting to obtain $F_{\nu}=4.36\pm0.28$ mJy, $6.55\pm0.47$ mJy, and $6.97\pm0.42$ mJy at central frequencies of 297 GHz, 349 GHz, and 357 GHz, respectively, providing a spectral index $\alpha=2.56\pm0.45$. We used this spectral index to scale all the continuum visibilities to a central frequency $\nu=345$ GHz for analysis of the coma dust mass and nucleus size. Figure~\ref{fig:contVis} shows that the visibilities were clearly consistent with a spatially extended source.

We followed the methods of \cite{Lellouch2022,Roth2025} to determine nucleus and dust contributions to the thermal continuum emission, performing least-squares fitting of coma models against the data. For a coma varying with nucleocentric distance as $r^{-1}$, the visibility amplitude varies with $uv$-radius ($\rho$) as $V(\rho) = \frac{Kc}{\nu\rho}$, where $K$ is a constant accounting for the emission properties of the dust. We find a best-fit coma model with $K=100\pm9$ Jy and a $(3\sigma)$ upper limit on the nucleus flux (taken to be a point source) $F(\nu=345 \mathrm{GHz})<1.7\times10^{-1}$ mJy. 

We assumed a geometric albedo of 5\%, a bolometric emissivity of 0.9, a radio emissivity ranging from 0.6--0.8, and a beaming parameter ranging from 0.9--1.2 as parameters for the Near Earth Asteroid Thermal Model \citep[NEATM;][]{Harris1998,Delbo2002} when deriving the nucleus size \citep{Lellouch2022}. For deriving the dust mass, we used opacities $2-4\times10^{-2}$ m$^2$ kg$^{-1}$, representative of cometary dust optical properties at 0.8 mm \citep{Boissier2012}. We find coma dust masses ranging from $M=(1.2-2.4)\times10^{11}$ kg, and (3$\sigma$) upper limits on the nucleus diameter $d<6.6-7.4$ km, where our nucleus value is consistent with results from JWST \citep[$d<8.4$ km;][]{Woodward2025}. We used this upper limit on nucleus size for setting the minimum radius of our coma radiative transfer models when analyzing molecular emission.

\subsection{Radiative Transfer Modeling}\label{subsec:radiative}
We modeled molecular line emission using the SUBLIME three-dimensional radiative transfer code for cometary atmospheres \citep{Cordiner2022}. SUBLIMED includes a full non-LTE treatment of coma gases, collisions with H$_2$O and electrons, and pumping by solar radiation, along with a time-dependent integration of the energy level population equations. We used explicitly calculated HCN-H$_2$O \citep{Zoltowski2025} and CO-H$_2$O \citep{Faure2020} collisional rates. We are unaware of published H$_2$CO-, CS-, or CH$_3$OH-H$_2$O collisional rates, so we calculated them using the thermalization approximation \citep{Crovisier1987,Biver1999,Bockelee-Morvan2012}, using an average collisional cross-section with H$_2$O of $5\times10^{-14}$ cm$^{-2}$. Photodissociation rates were adopted from \cite{Hrodmarsson2023}. We scaled the $Q$(H$_2$O) measured in \cite{Ejeta2025} at \rh{} = 2.35 au $(3.65\pm0.66\times10^{28}$ s$^{-1}$) to the \rh{} = 2.1 au of the ALMA observations assuming insolation-driven (\rh{}$^{-2}$) dependence. We used the resulting $Q$(H$_2$O) = $4.65\times10^{28}$ s$^{-1}$ for setting the activity level and density of our coma radiative transfer models.

\begin{deluxetable*}{cccccccccc}
\tablenum{3}
\tablecaption{Molecular Kinematics and Composition of C/2017 K2\label{tab:kinematics}}
\tablewidth{0pt}
\tablehead{
\colhead{Species} & \colhead{$v_1^{(a)}$} & \colhead{$v_2^{(b)}$} & \colhead{$Q_1/Q_2^{(c)}$} & \colhead{$\gamma^{(d)}$} & \colhead{$\phi^{(e)}$} & \colhead{$\psi^{(f)}$} & \colhead{$L_p^{(g)}$} & \colhead{$Q_x/Q_\mathrm{H2O}^{(h)}$} &  \colhead{$Q_x/Q_\mathrm{CO}^{(i)}$} \\
\colhead{} & \colhead{(\kms{})} & \colhead{(\kms{})} & \colhead{} & \colhead{($\degr$)} & \colhead{($\degr$)} & \colhead{($\degr$)} & \colhead{(km)} & \colhead{(\%)} & \colhead{(\%)}
}
\startdata
\multicolumn{10}{c}{2022 Sept. 21-24, \rh{}=2.11 au, $\Delta$=2.35 au, $\phi_\mathrm{STO}=25\degr$, $\psi_{\sun}=106\degr$} \\
\hline
CO & $0.40\pm0.02$ & $0.39\pm0.01$ & $0.52\pm0.06$ & $92\pm1$ & $22\pm2$ & $121\pm4$ & $182^{+60}_{-45}$ & $7.3\pm0.3$ & 100 \\
HCN & $0.410\pm0.001$ & $0.382\pm0.001$ & $1.15\pm0.02$ & $91\pm14$ & $27\pm1$ & $108\pm1$ & $249\pm6$ & $0.152\pm0.007$ & $2.1\pm0.1$ \\ 
CH$_3$OH & $0.452\pm0.003$ & $0.362\pm0.004$ & $2.12\pm0.06$ & $93\pm1$ & $24\pm1$ & $120\pm1$ & $126^{+19}_{-16}$ & $4.53\pm0.05$ & $62\pm3$ \\
CS & $0.43\pm0.01$ & $0.44\pm0.02$ & $1.5\pm0.2$ & $104\pm4$ & $15\pm3$ & $91\pm6$ & $813^{+160}_{-134}$ & $0.047\pm0.003$ & $0.64\pm0.05$ \\
$o-$H$_2$CO & $0.44\pm0.02$ & $0.37\pm0.02$ & $2.1\pm0.3$ & $106\pm2$ & $7\pm2$ & $125\pm5$ & $2722^{+695}_{-594}$ & $0.08\pm0.01$ & $1.1\pm0.2$ \\
$p-$H$_2$CO & (0.44) & (0.37) & (2.1) & (106) & (7) & (125) & (2722) & $0.028\pm0.003$ & $0.38\pm0.04$ \\
\hline
\hline
\enddata
\tablecomments{\sups{a} Gas expansion speed in the sunward jet $(R_1)$. \sups{b} Gas expansion speed in the anti-sunward region $(R_2)$. \sups{c} Ratio of production rates in the sunward vs.\ anti-sunward regions $(R_1/R_2)$. \sups{d} Half-opening angle of the sunward jet. \sups{e} Phase angle of the jet. \sups{f} Position angle of the sunward jet. \sups{g} Best-fit parent scale length from visibility modeling. \sups{h} Molecular abundance with respect to H$_2$O. \sups{i} Molecular abundance with respect to CO.
}
\end{deluxetable*}

\subsubsection{Determination of Outgassing Geometry}\label{subsubsec:kinematics}
Many of the spectral lines in K2 are clearly asymmetric, ruling out an isotropic coma model.  We followed methods applied in previous comet studies \citep{Cordiner2022,Roth2023,Roth2025}, dividing the coma into two outgassing regions, $R_1$ and $R_2$, each with independent molecular production rates ($Q_1$,$Q_2$), gas expansion velocities ($v_1$,$v_2$), and parent scale lengths ($L_{p1},L_{p2}$). $L_p$ parametrizes the distance from the nucleus at which a molecule formed ($L_p$ = 0 km is direct nucleus release) and is related to the parent photodissociation rate as $L_p = v/\beta_p$. 

Region $R_1$ is defined as the conical region with half-opening angle $\gamma$, originating at the nucleus surface and oriented at a phase angle $\phi$ with respect to the observer and position angle $\psi$ in the plane of the sky, and region $R_2$ the remainder of the coma \citep[see Fig. 8 of][]{Cordiner2023}. We initially assumed a constant coma kinetic temperature with radius and determined the spatial distribution of each species in $R_1, R_2$ using a Haser formalism \citep{Haser1957}:
\begin{equation}
    n_d(r) = \frac{Q_i}{4 \pi v_i r^2}\frac{\frac{v_i}{\beta_d}}{\frac{v_i}{\beta_p}-\frac{v_i}{\beta_d}}\left[\exp{\left(-\frac{\beta_p}{v_i}r\right)}-\exp{\left(-\frac{\beta_d}{v_i}r\right)}\right]
    \vspace{0.1cm}
\end{equation}
\noindent where $Q_i$ and $v_i$ are the molecular production rate (s$^{-1}$) and gas expansion velocity (km\,s$^{-1}$) in each coma region, and $\beta_p, \beta_d$ are the parent and daughter molecular photodissociation rates (s$^{-1}$), respectively. 

We let $Q_1/Q_2, v_1, v_2, \gamma, \phi, \psi,$ and the abundance vary as free parameters. $Q_1/Q_2, \gamma, \phi$, and $\psi$ were initialized as 1.0, 90$\degr$, $\phi_{\mathrm{STO}}$, and $\psi_{\sun}$. Our results are summarized for each molecule in Table~\ref{tab:kinematics}. Our results are consistent with a roughly symmetric split into hemispheric outgassing ($\gamma\sim90\degr$); however, there is variation in the orientation of the sunward-facing region, with H$_2$CO showing the most significant deviation ($\phi=7\pm2\degr, \psi=125\pm5\degr$) from the projected Sun-comet line ($\phi_{\mathrm{STO}}=25\degr, \psi_{\sun}=106\degr$). These results highlight the capability of interferometers such as ALMA to reveal the complex outgassing configurations present in cometary comae.

\begin{figure}
\plotone{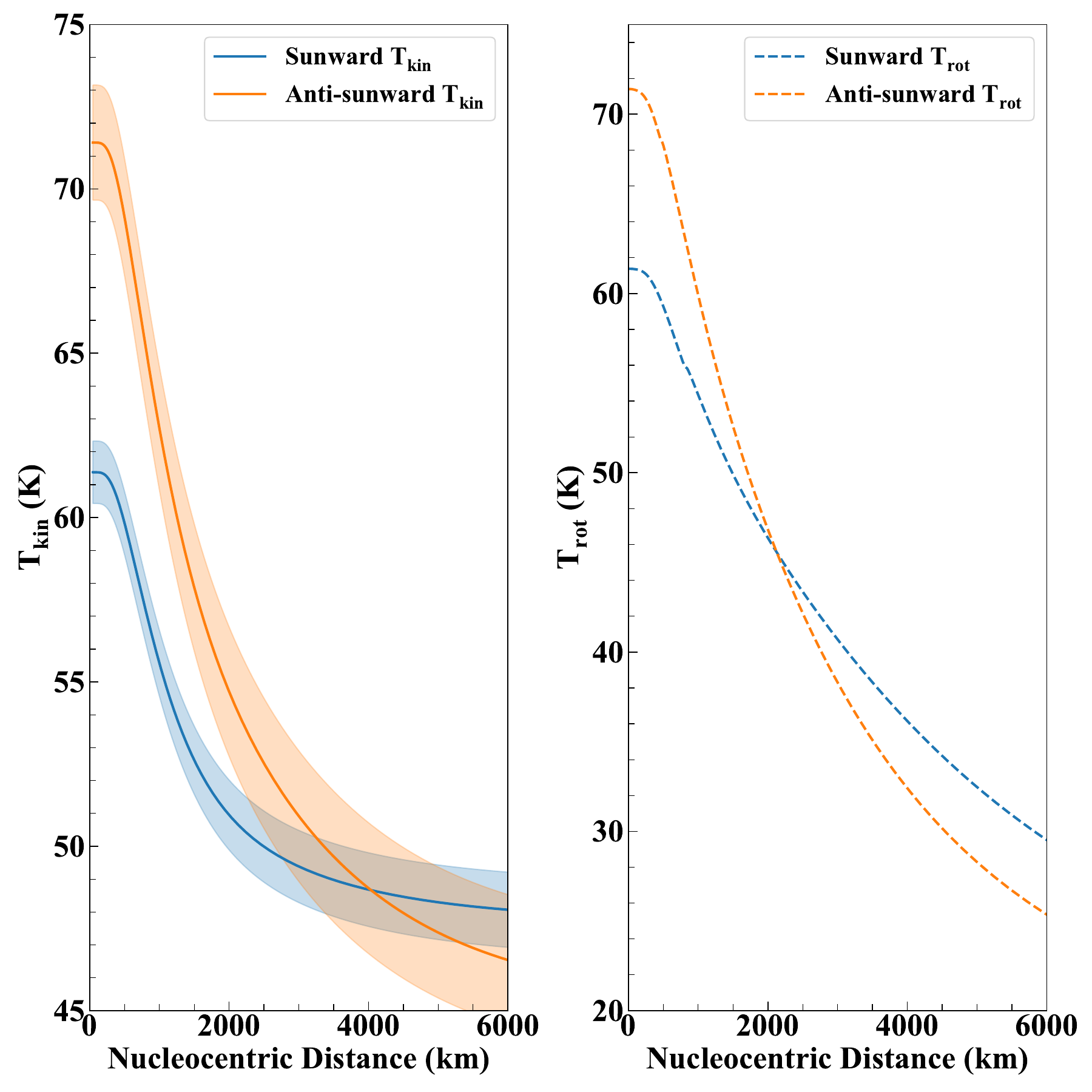}
\caption{\textbf{Left.} Sunward and anti-sunward best-fit radial kinetic temperature profiles for CH$_3$OH in C/2017 K2. The shaded regions show the $1\sigma$ uncertainties. \textbf{Right.} Sunward and anti-sunward radial rotational temperature profiles calculated from the best-fit SUBLIME model rotational populations. 
\label{fig:tprofile}}
\end{figure}

\subsubsection{Determination of Radial Temperature Profile}\label{subsubsec:tprofile}
The radial temperature profile $(T(r))$ of the coma was then determined by fitting all CH$_3$OH transitions simultaneously. Following \cite{Cordiner2023}, we used a smoothed, segmented, linear $T_i(r)$ function (in $\log(r)-T$ space where $r\in[\theta_\mathrm{min},\theta_\mathrm{MRS}]$ for each region $R_i$, using a variable number of segments $(n)$ of equal length $(l_s)$ in $\log(r)$ space, where $\theta_\mathrm{MRS}$ is the maximum recoverable scale. Smoothing was performed (in $\log(r)$ space) with a Gaussian of FWHM equal to $l_s$. We incrementally increased the model complexity from the simplest assumption of a constant temperature until we obtained an acceptable fit to the data. This approach minimizes the number of model free parameters while ensuring sufficient degrees of freedom to reproduce the data. We avoided ill-conditioned models (with too many segments) by testing for large-amplitude ripples in $T(r)$. 

We set the temperature at a constant value interior to a radius equal to half the minor axis of the synthesized beam  $(r=\theta_\mathrm{min}/2)$ and exterior to a radius equal to half the maximum recoverable scale $(r=\theta_\mathrm{MRS}/2)$. A good fit was obtained for $n=3$ variable points in $T_i(r)$. The best-fit points are $T_1 = [60.0\pm1.0$ K, $52.3\pm1.0$ K, $48.3\pm1.1$ K] and $T_2 = [69.4\pm1.8$ K, $57.4\pm1.9$ K, $47.4\pm2.0$ K] at radial points $r=[480$ km, 1550 km, 5000 km]. Figure~\ref{fig:tprofile} shows the best-fit three-segment temperature profile for CH$_3$OH in K2. A line-by-line model-data comparison is shown in Appendix~\ref{sec:spectra}.

\subsubsection{Determination of Molecular Production Mechanisms}\label{subsubsec:parentScale}
Finally, we worked to discern the modes of molecular production in K2. Coma volatiles were identified as ``parent'' or ``primary'' (produced via direct sublimation from the nucleus and thus directly indicative of its composition), as ``product'' or ``daughter'' (produced by photolysis of gas-phase species in the coma), or as an extended source \citep[these cannot be explained solely by gas-phase processes and are likely formed from photo and thermal degradation of dust;][]{Cottin2008}. 
By comparing our derived $\beta_p$ against those of molecules with compatible photolysis pathways (e.g., CS$_2$ to CS), we can quantitatively test the identity of progenitor materials in the coma. If no compatible gas-phase precursor is known, we assume the molecule was produced by the breakdown of an unknown refractory compound in the coma.

\begin{figure*}
\gridline{\fig{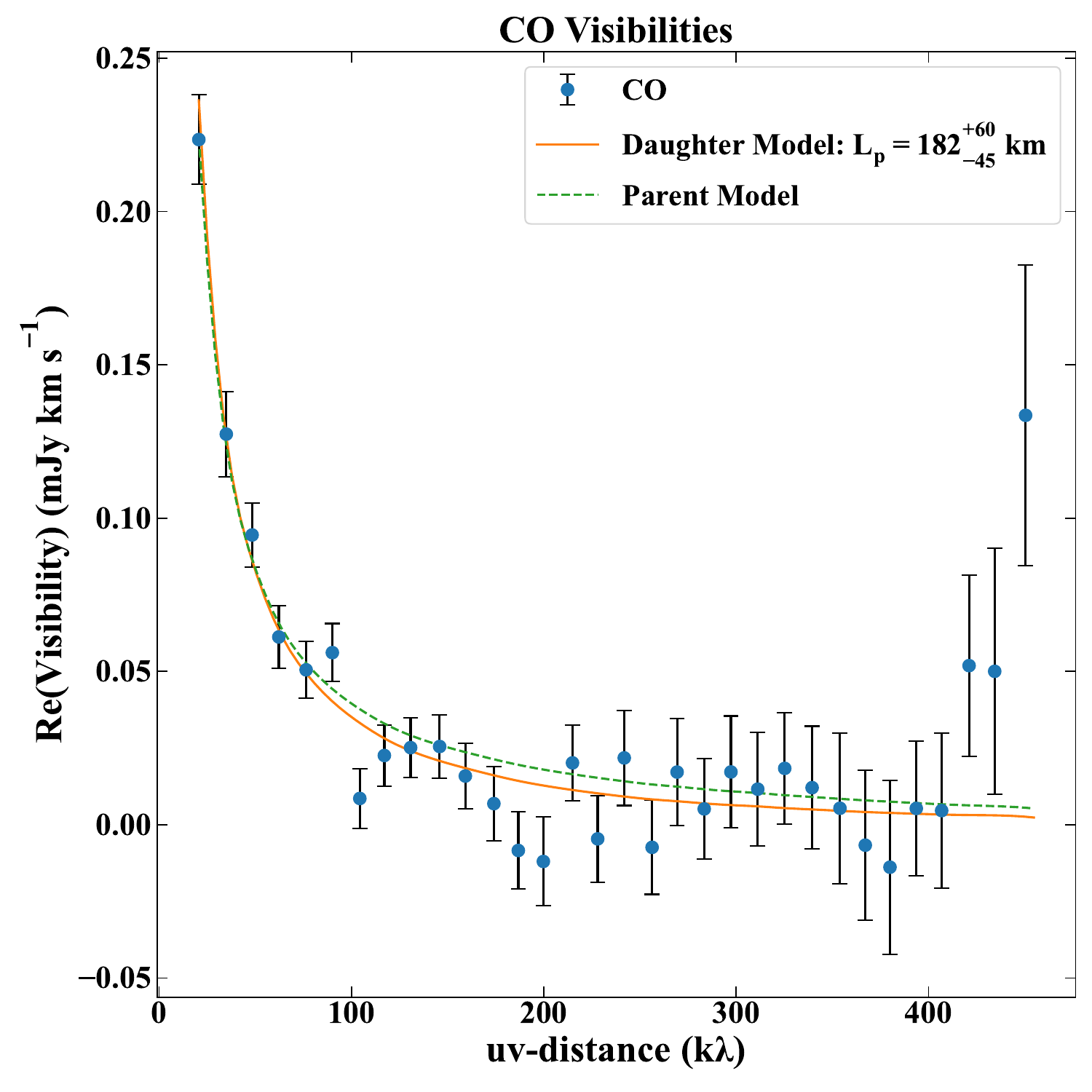}{0.3\textwidth}{(A)}
          \fig{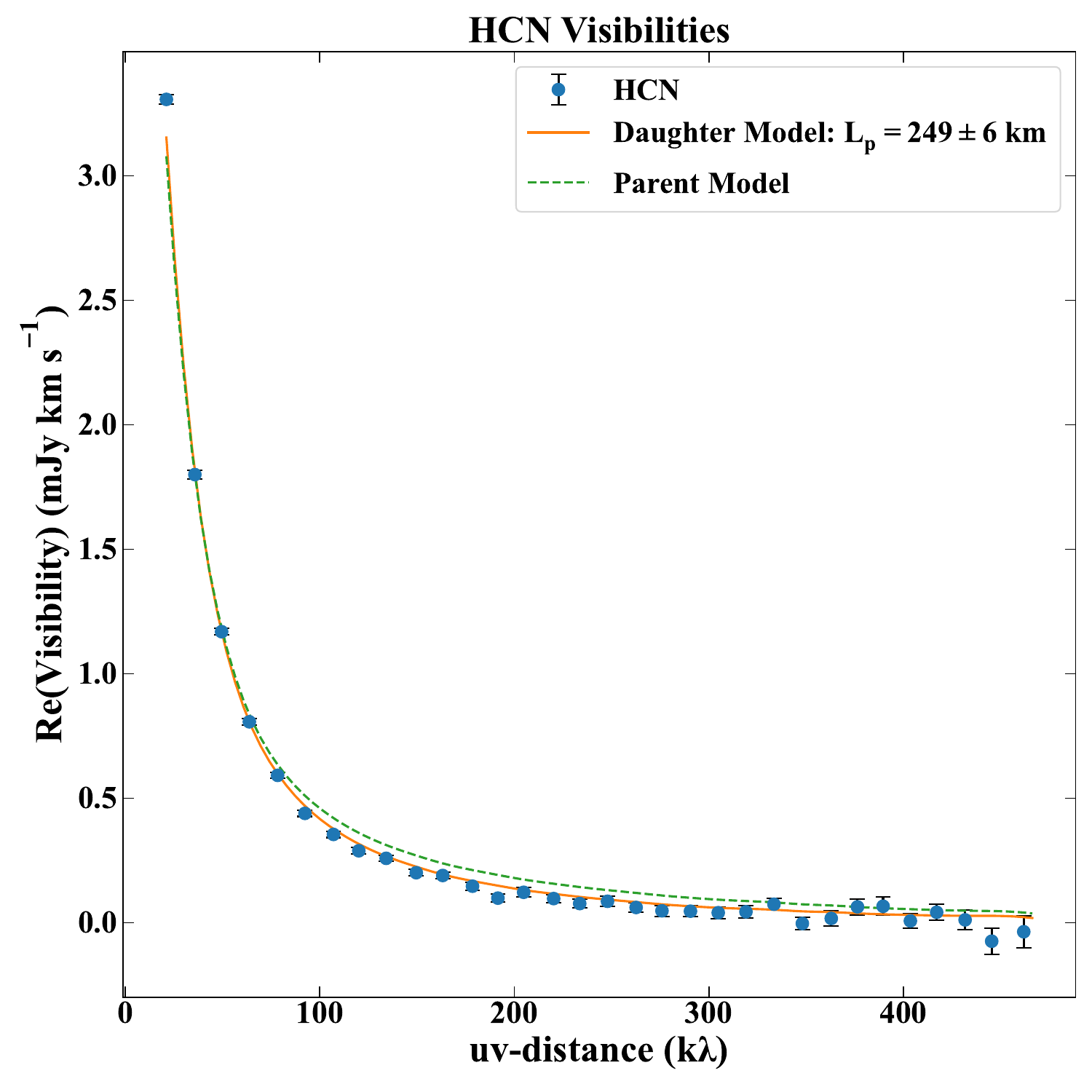}{0.3\textwidth}{(B)}
          \fig{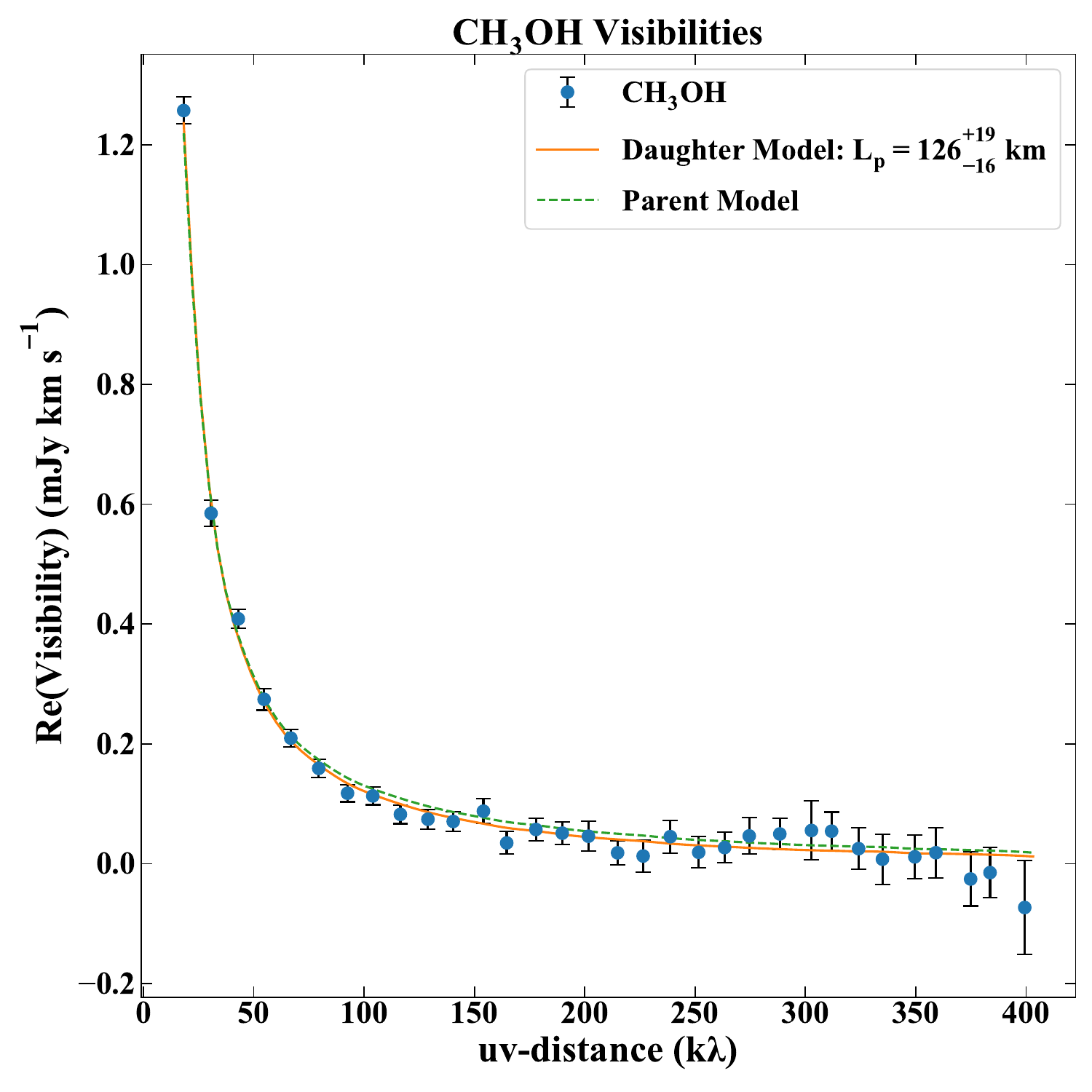}{0.3\textwidth}{(B)}
}
\gridline{\fig{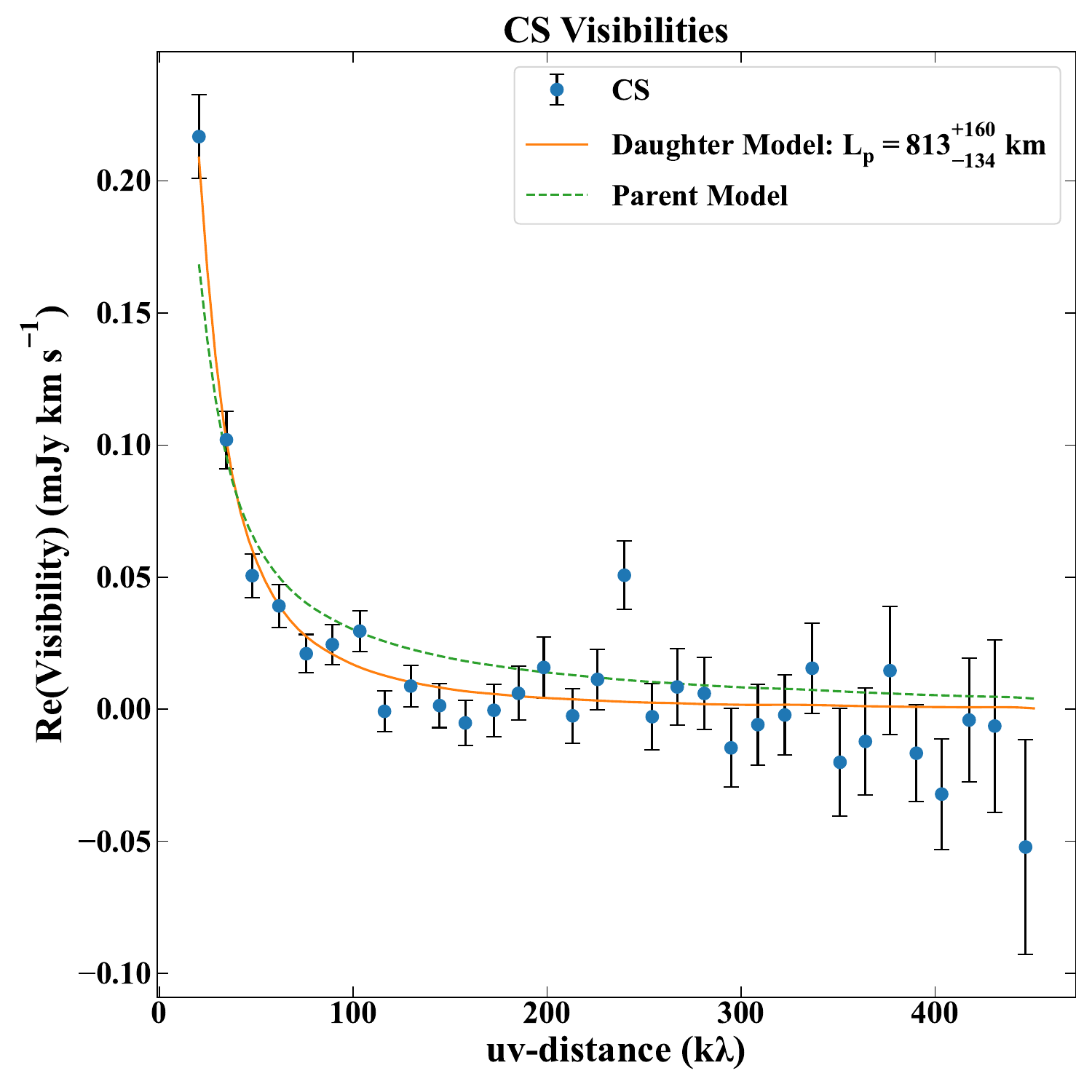}{0.3\textwidth}{(C)}
          \fig{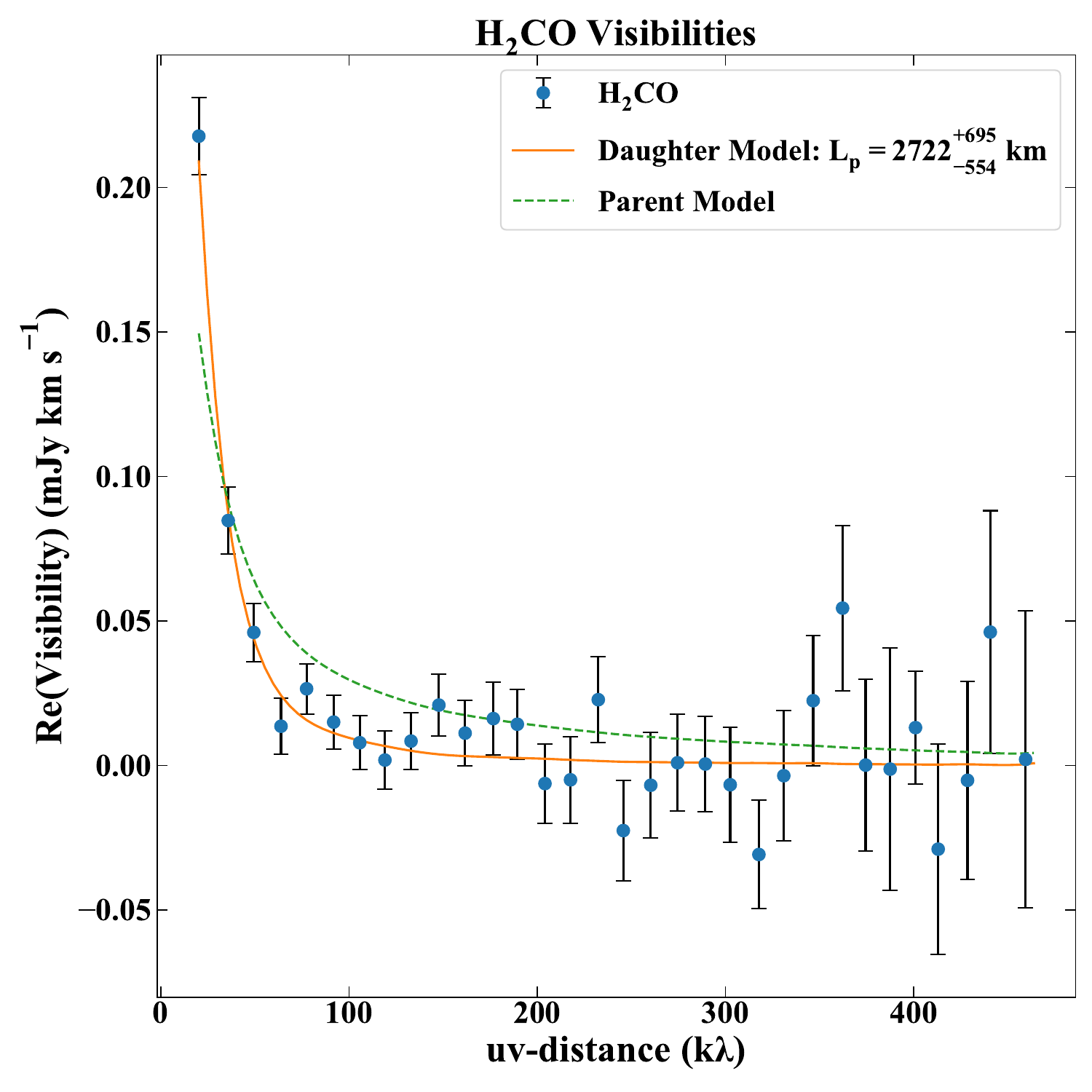}{0.3\textwidth}{(D)}
}
\caption{\textbf{(A)--(D).} Real part of the observed visibility amplitude as a function of projected baseline length for CO, HCN, CH$_3$OH, CS, and H$_2$CO. Best-fit daughter models, as well as a parent model, are overplotted for comparison. The $uv$ distance has been plotted in units of k$\lambda$.
\label{fig:visCurves}}
\end{figure*}

The density of coma parent molecules peaks at the nucleus before falling off exponentially with increasing nucleocentric distance owing to adiabatic expansion and photolysis. To adequately sample the often asymmetric (and nonlinear) uncertainties on $L_p$ we followed the methods of \cite{Cordiner2023}, generating the $\chi^2$ surface for each species as a function of $L_p$. We iterated over a range of $L_p$ values, performing least squares fitting where the production rate and expansion speed were allowed to vary while holding all other parameters fixed. We recorded the $\chi^2$ statistic of the optimal SUBLIME model (calculated as the goodness of fit for the observed vs.\ model visibilities) found for each fixed $L_p$ value. We then examined the $\Delta\chi^2$ curve generated using cubic spline interpolation between each value of $L_p$, where $\Delta\chi^2(L_p) = \chi^2(L_p) - \chi^2_{min}$. We obtained the 1$\sigma$ and 2.6$\sigma$ uncertainties from the values for $\Delta\chi^2$ = 1 (68\% confidence) and 6.63 (99\% confidence) thresholds. Figures~\ref{fig:visCurves} shows our observed and best-fit model visibility curves. In general, the visibility amplitudes are indistinguishable from zero at baselines longer than 150 k$\lambda$. Although the CO visibilities appear to show a rise on the very longest baselines, the uncertainties are large and they are still consistent with zero at the $2\sigma-3\sigma$ level.  Appendix~\ref{sec:spectra} includes a model-data comparison of our spectra as a function of baseline.

\subsection{Ortho-to-Para Ratio of H$_2$CO}\label{subsec:h2co}
We detected one ortho- and four para-H$_2$CO transitions in K2 (Table~\ref{tab:lines}). Of these, the ortho $J_{Ka,Kc} = 5_{1,5}-4_{1,4}$ and para $5_{0,5}-4_{0,4}$ transitions were measured simultaneously, enabling a sensitive measure of the ortho-to-para ratio (OPR). The detected ortho and para transitions have a small spread in $E_u$, limiting uncertainties relative to radiative transfer modeling. 

We performed our initial visibility modeling on the stronger ortho transition, determining $L_p$, the outgassing kinematics, and the ortho abundance. We then fixed $Q_1/Q_2, v_1, v_2, \gamma, \phi, \psi,$ and $L_p$ and let the daughter abundance and ortho-to-para ratio vary as free parameters while fitting ortho and para transitions simultaneously. We first fit only the (simultaneously measured) ortho $J_{Ka,Kc} = 5_{1,5}-4_{1,4}$ and para $5_{0,5}-4_{0,4}$ transitions from the 357 GHz correlator setting, finding OPR=$3.0\pm0.5$. 

We then included the $J_{Ka,Kc} = 4_{0,4}-3_{0,3}, 4_{2,3}-3_{2,2}$, and $4_{2,2}-3_{2,1}$ transitions from the 297 GHz correlator setting. This improved statistics for the weaker para transitions at the cost of exposure to potential short-term variability in comet activity. The resulting OPR is $2.9\pm0.4$, in agreement with our first iteration. We adopt the latter value for subsequent discussion.

\section{Discussion and Interpretation} \label{sec:discussion}

\subsection{Modes of Molecular Production in C/2017 K2}\label{subsec:parentScales}
Coupling the capabilities of state-of-the-art interferometers with sophisticated radiative transfer modeling provides quantitative measures of the parent scale length, $L_p$, and parent photodissociation rate, $\beta_p$, which in turn enables investigations of progenitor materials for molecules with significant production in the coma. 

To test whether a molecule is compatible with synthesis from gas-phase chemistry in the coma, it is convenient to scale the derived $\beta_p$ to \rh{}=1 au assuming insolation-driven dependence (i.e., \brh{}=$\beta$(\rh{})\rh{}$^2$). This enables a straightforward comparison against solar photodissociation rates of molecules with compatible photolysis pathways \citep{Hrodmarsson2023}. For molecules likely to be produced from thermal or photo degradation of grains \citep[e.g., H$_2$CO;][]{Biver1999} it is more useful to directly compare $L_p$, as the dust does not necessarily move at the same speed as the gas, nor does its degradation rate follow a simple \rh{}$^{-2}$ dependence. Table~\ref{tab:beta} shows \brh{} and $L_p$ for HCN, CS, H$_2$CO, and CH$_3$OH in K2 for comparison against their measure in other comets with ALMA.

\begin{deluxetable*}{cccccc}
\tablenum{4}
\tablecaption{Properties of Progenitor Material in Comets Studied with ALMA \label{tab:beta}}
\tablewidth{0pt}
\tablehead{
\colhead{Molecule} & \colhead{K2} & \colhead{46P/Wirtanen} & \colhead{C/2015 ER61} & \colhead{C/2012 S1 (ISON)} & \colhead{C/2012 F6 (Lemmon)} \\
\colhead{\rh{}} & \colhead{(2.11 au)} & \colhead{(1.07 au)} & \colhead{(1.14 au)} & \colhead{(0.54 au)} & \colhead{(1.47 au)}
}
\startdata
\multicolumn{6}{c}{Parent Photodissociation Rates Scaled to \rh{} = 1 au (s$^{-1}$)} \\
HCN & $(7.33^{+0.18}_{-0.17})\times10^{-3}$ & $>2.47\times10^{-1}$& ($1.55^{+2.34}_{-0.78}$) $\times$ $10^{-2}$& $>5.8$ $\times$ $10^{-3}$ & $>3.05$ $\times$ $10^{-2}$ \\
CS & $(2.36^{+0.46}_{-0.39})\times10^{-3}$ & $(1.12^{+0.25}_{-0.21})\times10^{-3}$ &($3.57^{+5.36}_{-2.17})$ $\times$ $10^{-4}$ & ($1.46^{+0.47}_{-0.29})$ $\times$ $10^{-3}$ & $\cdot \cdot \cdot$ \\
H$_2$CO & $(7.20^{+2.01}_{-1.46})\times10^{-4}$ & $(8.50^{+2.12}_{-1.89})\times10^{-4}$ & ($3.08^{+1.76}_{-1.03})$ $\times$ $10^{-4}$ & ($1.04^{+0.23}_{-0.16})$ $\times$ $10^{-3}$ & ($1.26^{+0.63}_{-0.63})$ $\times$ $10^{-3}$\\
CH$_3$OH & $(1.59^{+0.32}_{-0.24})\times10^{-2}$ & $(2.32^{+0.56}_{-0.38})\times10^{-2}$ & $(2.34^{+15.2}_{-1.92})\times10^{-3}$ & ... & ... \\
\hline
\multicolumn{6}{c}{Parent Scale Lengths (km)} \\
HCN & $249\pm6$ & $<3$ & $50^{+50}_{-30}$ & $<50$ & $<50$ \\
CS & $813^{+160}_{-134}$ & $675^{+159}_{-124}$ & $2000^{+3100}_{-1200}$ & $200\pm50$ & ... \\
H$_2$CO & $2722^{+695}_{-594}$ & $876^{+250}_{-175}$ & $2200^{+1100}_{-800}$ & $280\pm50$ & $1200^{+1200}_{-400}$ \\
CH$_3$OH & $126^{+19}_{-16}$ & $36\pm7$ & $300^{+1400}_{-260}$ & ... & ...
\enddata
\tablecomments{Photodissociation rate (s$^{-1}$) corrected for \textit{v}\subs{exp} and \textit{r}\subs{H} at the time of the observations. Rates for 46P are from \cite{Cordiner2023}, whereas rates for ER61 are from \cite{Roth2021a}. Rates for ISON and Lemmon are taken from \cite{Cordiner2014} except for CS \citep{Bogelund2017}.
}
\end{deluxetable*}

\subsubsection{CO, CH$_3$OH, and HCN}\label{subsubsec:co_lp} 
K2 is just the third comet with reported ALMA measurements of CO \citep[the others being 2I/Borisov at \rh{}=2.01 au and C/2014 UN271 measured at \rh{}=16.6 au;][]{Cordiner2020,Roth2025}. For CO in K2, $L_p=182^{+60}_{-45}$ km, considerably smaller than $L_p<2500$ km $(1\sigma)$ measured in C/2014 UN271. However, it is worth keeping in mind the significant differences in observing circumstances (coma excitation conditions, geocentric distance, angular resolution) for these two comets. No $L_p$ was reported for CO 2I/Borisov. The $L_p$ for CO in K2 corresponds to \brh{}=$(9.79^{+3.21}_{-2.43})\times10^{-3}$ \ps{}. In terms of cometary molecules with photolysis pathways to CO, H$_2$CO has the most similar photodissociation rate at \rh{} = 1 au ($\beta=2.03\times10^{-4}$ \ps{}), yet this rate is too slow for CO in K2 at the $\sim$4$\sigma$ level, and the H$_2$CO molecular abundance in K2 is orders of magnitude lower, ruling it out as a potential parent. 

CH$_3$OH and HCN have \brh{} and $L_p$ which are in formal agreement with CO, and similarly there are not known cometary molecules with photolysis pathways to either species whose photodissociation rates are compatible with those in K2. In the absence of gas-phase precursors, a potential alternative is production of these volatiles from icy grain sublimation. VLT/MUSE studies of  K2 at \rh{}=3.5 au found that millimeter-sized dust chunks were spatially co-located with OI$^1$D emission, suggesting that the dust chunks were coated with H$_2$O ice \citep{Kwon2023}. JWST spatial-spectral studies of K2 at \rh{} = 2.35 au at mid- and near-infrared wavelengths required H$_2$O production from icy grains to explain its coma spatial distribution \citep{Woodward2025}, and calculated an H$_2$O active fraction $>86\%$, indicating that K2 is a hyperactive comet.

CH$_3$OH production from icy grains has often been indicated in hyperactive comets. For instance, ALMA and Keck/NIRSPEC studies of CH$_3$OH in hyperactive comet 46P/Wirtanen were consistent with its production from icy grain release \citep{Bonev2021,Roth2021b,Cordiner2023}. Likewise, \cite{Drahus2012} found evidence for HCN and CH$_3$OH sublimation from icy grain sources in the coma of EPOXI target 103P/Hartley 2. Although extended sources of CO were suggested based on IR observations of C/1995 O1 (Hale-Bopp) \citep{DiSanti2003}, analysis of measurements with the Plateau de Bure interferometer found CO was instead produced by direct nucleus release \citep{Bockelee2010}. 

The liftimes for CH$_3$OH, CO, and HCN in K2 are $\tau_{\mathrm{CH3OH}}=279^{+42}_{-35}$ s, $\tau_{\mathrm{CO}}=455^{+150}_{-113}$ s, and $\tau_{\mathrm{HCN}}=607\pm15$ s, respectively. Based on calculations by \cite{Beer2006}, these would correspond to a populaton of sub-micron dirty ice grains, considerably smaller than the mm-sized icy grain chunk sizes found by \cite{Kwon2023}. The presence of icy grain CH$_3$OH sublimation in K2 may also consistent with our measured kinetic temperature profiles (\S~\ref{subsec:thermal}).

\subsection{CS}\label{subsubsec:cs}
The origin of CS in comets is a subject of debate. Observations of C/2015 ER61, C/2020 F3 (NEOWISE), 46P/Wirtanen, C/2021 A1 (Leonard), and C/2022 E3 (ZTF) showed that CS production could not be explained by CS$_2$ photolysis alone and required significant production from extended sources \citep{Roth2021a,Biver2022,Cordiner2023,Biver2024}. However, the CS$_2$ photodissociation rate at \rh{} = 1 au ($\beta=2.12\times10^{-3}$ \ps{}) is in excellent agreement with \brh{}$=(2.35^{+0.46}_{-0.39})\times10^{-3}$ \ps{} for CS in K2, indicating that it can be explained entirely by CS$_2$ photolysis.

\subsubsection{H$_2$CO}\label{subsubsec:ch2co}
On the other hand, H$_2$CO requires significant coma production to explain its spatial distribution. The derived $L_p$ for H$_2$CO in K2 is in formal agreement with those measured for C/2015 ER61 and C/2012 F6 (Lemmon), although both were observed at considerably smaller \rh{}.  Collectively, none of the H$_2$CO \brh{} are compatible with any known photolytical precursor to H$_2$CO in comets, and are instead consistent with the non-Haser behavior of the unknown (likely refractory) H$_2$CO progenitor \citep[e.g.][]{Roth2021a,Cordiner2023}.

\subsection{Thermal Structure of the Coma}\label{subsec:thermal}
Figure~\ref{fig:tprofile} shows the modeled kinetic and rotational temperature profiles in K2. Owing to the relatively compact array configuration during our observations, the ALMA measurements span a large range of nucleocentric distances. This includes from the inner coma \cite[$r<1000$ km;][]{Marschall2024} where collisions dominate and thermalize the rotational temperature to the kinetic temperature of the gas, to the more distant coma ($r\sim10,000$ km) where collisions become rarer and fluorescence plays an ever larger role in governing excitation with increasing nucleocentric distance. In terms of kinetic temperature, the anti-sunward hemisphere of the coma was $\sim$10 K warmer near the nucleus, yet the dayside cooled considerably quicker than the nightside. This hemispheric asymmetry is indicative of different balances of heating and cooling mechanisms in the day and night sides of the coma. A similar trend is seen for the rotational temperature, and its decoupling from the kinetic temperature at $r>400$ km can be understood in terms of the gradual departure from thermal equilibrium as collisional excitation continues to weaken at larger nucleocentric distances. However, the turnover between hemispheres occurs at smaller nucleocentric distances for the rotational temperature than for the kinetic temperature.

JWST observations of K2 noted sunward/anti-sunward asymmetries for the H$_2$O rotational temperature, which was attributed to the presence of icy-grain sublimation \citep{Woodward2025}. However, the dichotomy in day vs.\ night side measured by JWST was in the opposite sense to that found by ALMA. It is worth noting the JWST H$_2$O maps sampled the inner coma ($r<1500$ km), representing a fraction of the distances covered by ALMA. 

Hemispheric asymmetries in temperature were also found in the inner coma ($r<1500$ km) of comet 46P/Wirtanen, with a lower near-nucleus temperature and more rapid cooling in the sunward hemisphere. This was attributed to a combination of more efficient adiabatic cooling in the sunward hemisphere and the heating effects of icy grain sublimation in the anti-sunward side \citep{Cordiner2023,Biver2021}. 

Similar to 46P/Wirtanen, K2 is a hyperactive comet \citep{Woodward2025}, requiring coma sources of H$_2$O (such as icy grain sublimation) to explain its excessive $Q$(H$_2$O) relative to its nucleus size. Coupled with the $L_p$ found for CH$_3$OH, CO, and and HCN (\S~\ref{subsubsec:co_lp}) and lines of evidence for icy grains from other wavelengths \citep{Kwon2023,Woodward2025}, it is reasonable to assume that icy grain sublimation played a significant role in K2's coma. Like 46P, the higher anti-sunward temperature and the quicker cooling on the sunward side of K2's coma are consistent with the combined effects of efficient adiabatic cooling and heating from icy grains.

\subsection{Comparison to Other Measurements of C/2017 K2 and the Comet Population}\label{subsec:compareK2}

Given its large discovery distance and high activity, K2 garnered significant attention and was the subject of numerous studies as it approached the Sun. Most relevant to this work (conducted at \rh{} = 2.11 au) are JWST and NASA-IRTF studies taken at \rh{} = 2.35 au \citep{Woodward2025,Ejeta2025}. We can also compare our molecular abundances against compositional averages found for Oort cloud comets measured to date \citep{DelloRusso2016a,Biver2024}. 

\begin{figure}
\plotone{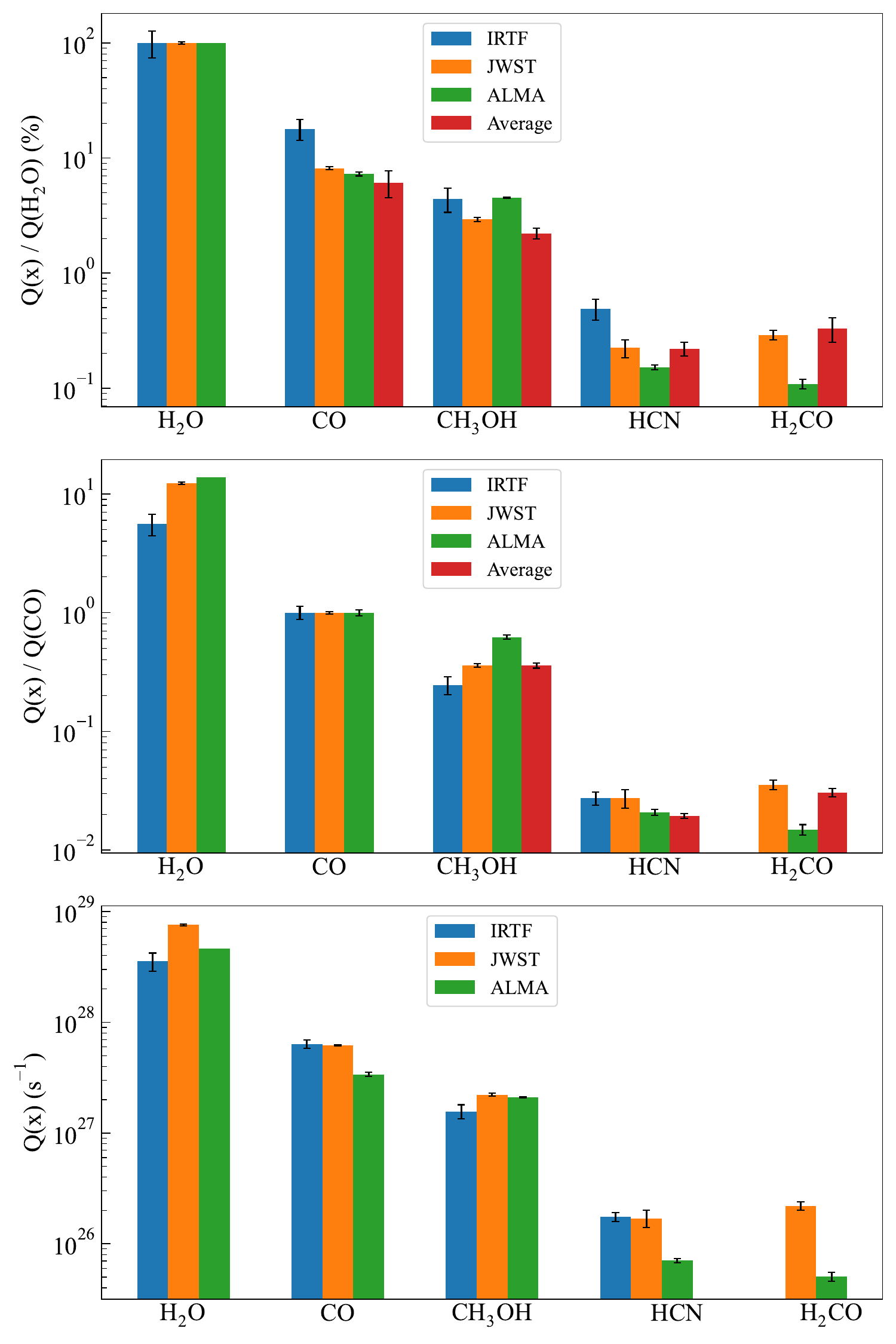}
\caption{\textbf{Left.} Molecular abundances in C/2017 K2  with respect H$_2$O measured with IRTF \citep{Ejeta2025}, JWST \citep{Woodward2025}, and ALMA (this work). \textbf{Middle.} As in the left panel, but abundances with respect to CO. \textbf{Right.} Absolute molecular production rates measured with each facility. Mean values are taken from \cite{DelloRusso2016a,Roth2023} and references therein.
\label{fig:k2compare}}
\end{figure}

\cite{Woodward2025} found $Q$(H$_2$O) roughly twice as high as \cite{Ejeta2025}: $(7.65\pm0.15)\times10^{28}$ s$^{-1}$ and $(3.65\pm0.66)\times10^{28}$ s$^{-1}$, respectively. These two studies reported molecular production rates assuming direct nucleus release, whereas our modeling can account for coma production. These subtleties are important whenever comparing molecular abundances. To disentangle the potential effects of short-term variability in production rates as well as the choice of reference molecule for calculating abundances, we compared absolute $Q_x$, $Q_x/Q_{CO}$, and $Q_x$/$Q_{H2O}$. Figure~\ref{fig:k2compare} shows the comparison.

For CO, our $Q$(CO) is lower than both JWST and IRTF, whereas our CO/H$_2$O is consistent with JWST (but lower than IRTF). Our Q(CH$_3$OH) is consistent within $2\sigma$ for JWST and IRTF, whereas our CH$_3$OH/H$_2$O is consistent with IRTF but higher than JWST, and our CH3OH/CO is higher than both facilities. HCN and H$_2$CO require more nuanced comparisons. Abundances measured in a given comet are often lower at radio wavelengths compared to near-IR observations for reasons unknown \citep{Biver2024,DelloRusso2016a}, yet our HCN/CO is consistent with the population mean. On the other hand, the association with distributed source production for H$_2$CO at radio wavelengths makes comparisons with infrared models assuming direct release fraught. However, our H$_2$CO/H$_2$O and H$_2$CO/CO abundances are both depleted compared to their relative population means.

The potential for short-term variation in absolute production rates notwithstanding, among the species measured with ALMA, JWST, and/or IRTF, the results indicate that K2 was average-to-enriched in CO, enriched in CH$_3$OH, average in HCN (for radio statistics), and depleted in H$_2$CO. Finally, CS was strongly strongly depleted in K2 compared to the range measured among comets \citep[0.03--0.2\% with respect to H$_2$O][]{Roth2021a}. This ``mixed'' composition (enriched in some volatiles, average in others, and depleted in the rest) has been observed in comets across dynamical families \citep[e.g.,][]{Roth2017,DiSanti2017,Cordiner2023}.

Likewise, such a mixed composition presents challenges for interpreting potential formation times and/or locations in the protoplanetary disk. \cite{Willacy2022} compared molecular abundances in a survey of comets measured at infrared wavelengths against predicted disk midplane ice-phase abundances as a function of time and radius, finding that no single combination of the two could reproduce the range of abundances measured in comets. The compositional catalog of K2 presents similar challenges: its HCN/H$_2$O abundance requires formation between 7-9 au any time within the first $10^8$ years of the disk. Although there are compatible combinations of time and location for its H$_2$CO/H$_2$O and CH$_3$OH/H$_2$O abundances, it is irreconcilable with any combination of disk time and location capable of reproducing its CO/H$_2$O abundance (see Figure 5 of \cite{Willacy2022}). Clearly, additional synergistic work between comet compositional studies and protoplanetary disk models are needed to discern potential formation pathways for individual comets.

\subsection{Ortho-to-Para Ratio and Spin Temperatures in Comets}\label{subsec:ortho-discuss}
Molecules with multiple hydrogen atoms can exist in distinct isomeric forms depending on the relative alignments of their nuclear spin states. Relevant to comets, molecules such as H$_2$O, NH$_2$, and H$_2$CO can exist in ortho (hydrogen spins parallel) or para (spin states antiparallel) forms. In our case, the rotational lines of ortho- and para-H$_2$CO are easily distinguished with facilities such as ALMA. 

The OPR of cometary molecules has been suggested as a ``cosmogonic indicator'' which may reveal information about the formation temperatures of cometary ices \citep{MummaCharnley}. The relative populations of the ortho- and para- forms can be related to a nuclear spin temperature ($T_\mathrm{spin}$), and nuclear spin state conversion via radiative or collisional processes is forbidden in the gas-phase. Hence, this $T_\mathrm{spin}$ has been previously interpreted as the formation temperature for cometary ices. The most commonly described cometary OPR is for H$_2$O \citep[e.g.,][]{Faggi2018}, whereas the sole OPR reported for cometary H$_2$CO is from 103P/Hartley 2 \citep[$2.12\pm0.59$, $T_{\mathrm{spin}}=13.5^{+6.7}_{-3.3}$ K;][]{Gicquel2014}. Our OPR of $2.9\pm0.4$ is in formal agreement with \cite{Gicquel2014} and implies $T_{\mathrm{spin}}>17$ K. It is tempting to conclude that this is consistent with the general picture of inheritance of interstellar ices into cometary nuclei. However, recent laboratory work found no correlation between H$_2$CO formation temperature and OPR \citep{Yocum2023}, with similar results found for H$_2$O \citep{Hama2018}, providing evidence that spin species statistics for molecules in cometary comae may not reflect cosmogonic values. Increasing the sample size of measured cometary OPR for H$_2$CO, coupled with laboratory work for spin ratios of additional cometary species such as CH$_3$OH and CH$_4$, is clearly needed to understand the significance of measured spin species statistics.
\section{Conclusion and Future Directions} \label{sec:conclusion}
The powerful ALMA telescope has opened new windows into the physics and chemistry of cometary comae over a range of heliocentric distances. Its large collecting area and high spectral and angular resolution are crucial for characterizing the faint, extended emission in these objects. Our study of comet C/2017 K2 (PanSTARRS) at \rh{} = 2.1 au adds to these works, revealing its pre-perihelion composition in an H$_2$O-dominated coma as measured at sub-millimeter wavelengths. Our measurements indicated that, compared to mean cometary values, K2 was enriched in CO and CH$_3$OH, average in HCN, and depleted in H$_2$CO and CS. CO, CH$_3$OH, and HCN were produced within $\sim$250 km of the nucleus and potentially experienced contributions from icy grain sublimation. H$_2$CO and CS were produced in the coma, with the former requiring production from an unknown progenitor in the extended coma and the latter consistent with production from CS$_2$ photolysis. The clumpy, asymmetric spatial distribution of H$_2$CO in particular shows a clear example of extended source production (Figures~\ref{fig:h2co-maps}, \ref{fig:allH2CO}). Spatially resolved measures of continuum emission indicated a coma dust mass of $1.2-2.4\times10^{11}$ kg and an upper limit on the nucleus diameter $d<6.6-7.4$ km, with the latter consistent with constraints made by JWST. 

These results will form the basis for upcoming reports of observations taken at larger pre-perihelion \rh{}, enabling a comparison of the molecular abundances, kinematics, and modes of production in K2 as it approached and crossed the H$_2$O sublimation zone, transitioning from CO- to H$_2$O-dominated outgassing. More broadly, our study demonstrates the capability for ALMA to perform serial studies to characterize cometary atmospheres across a range of \rh{} and coma conditions. Such studies will be increasingly commonplace as advanced facilities such as the Vera Rubin Observatory Legacy Survey of Space and Time (LSST) contribute to dramatically increased comet discovery rates, enabling us to characterize bright and distantly active comets such as C/2017 K2.

\begin{acknowledgments}
This work was supported by the NASA Solar System Observations program (80NSSC24K1324: N.X.R., S.N.M., M.A.C.), the Planetary Science Division Internal Scientist Funding Program through the Fundamental Laboratory Research (FLaRe) work package (N.X.R., S.N.M., M.A.C., S.B.C.), as well as the NASA Astrobiology Institute through the Goddard Center for Astrobiology (proposal 13-13NAI7-0032; S.N.M., M.A.C., S.B.C.). It makes use of the following ALMA data: ADS/JAO.ALMA \#2021.1.00862.S. ALMA is a partnership of ESO (representing its member states), NSF (USA) and NINS (Japan), together with NRC (Canada), NSTC and ASIAA (Taiwan), and KASI (Republic of Korea), in cooperation with the Republic of Chile. The Joint ALMA Observatory is operated by ESO, AUI/NRAO and NAOJ. The National Radio Astronomy Observatory is a facility of the National Science Foundation operated under cooperative agreement by Associated Universities, Inc. 
We thank two anonymous referees for their feedback, which we feel improved the manuscript.
\end{acknowledgments}

\software{Astropy \citep{astropy:2013, astropy:2018, astropy:2022},
Astroquery \citep{Ginsburg2019},
CASA \citep{McMullin2007},
lmfit \citep{Newville2016},
vis-sample \citep{Loomis2018}
}

\appendix
\section{H$_2$CO Spectral Gallery}\label{sec:h2co_imaging}
\begin{figure*}
\gridline{\fig{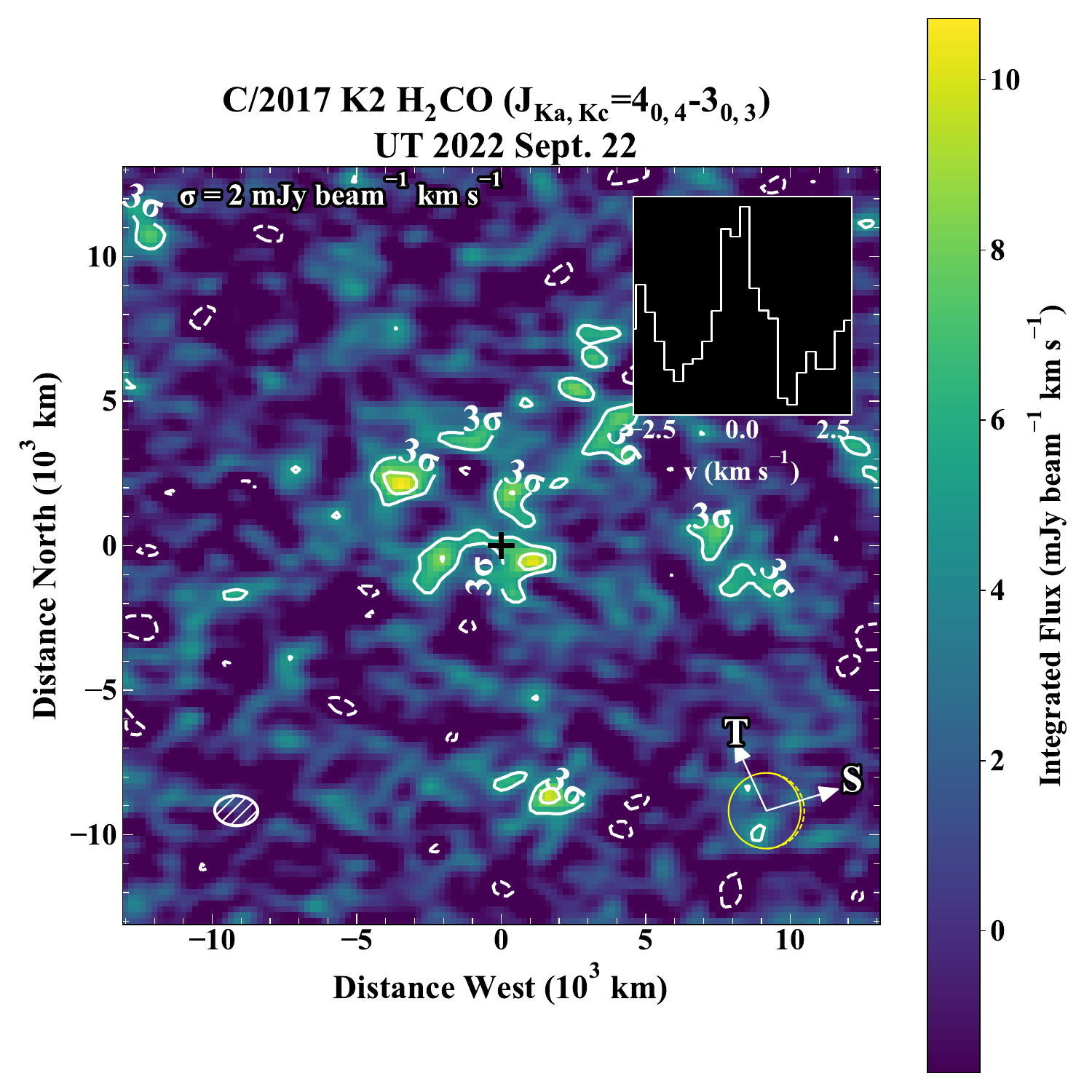}{0.3\textwidth}{(A)}
          \fig{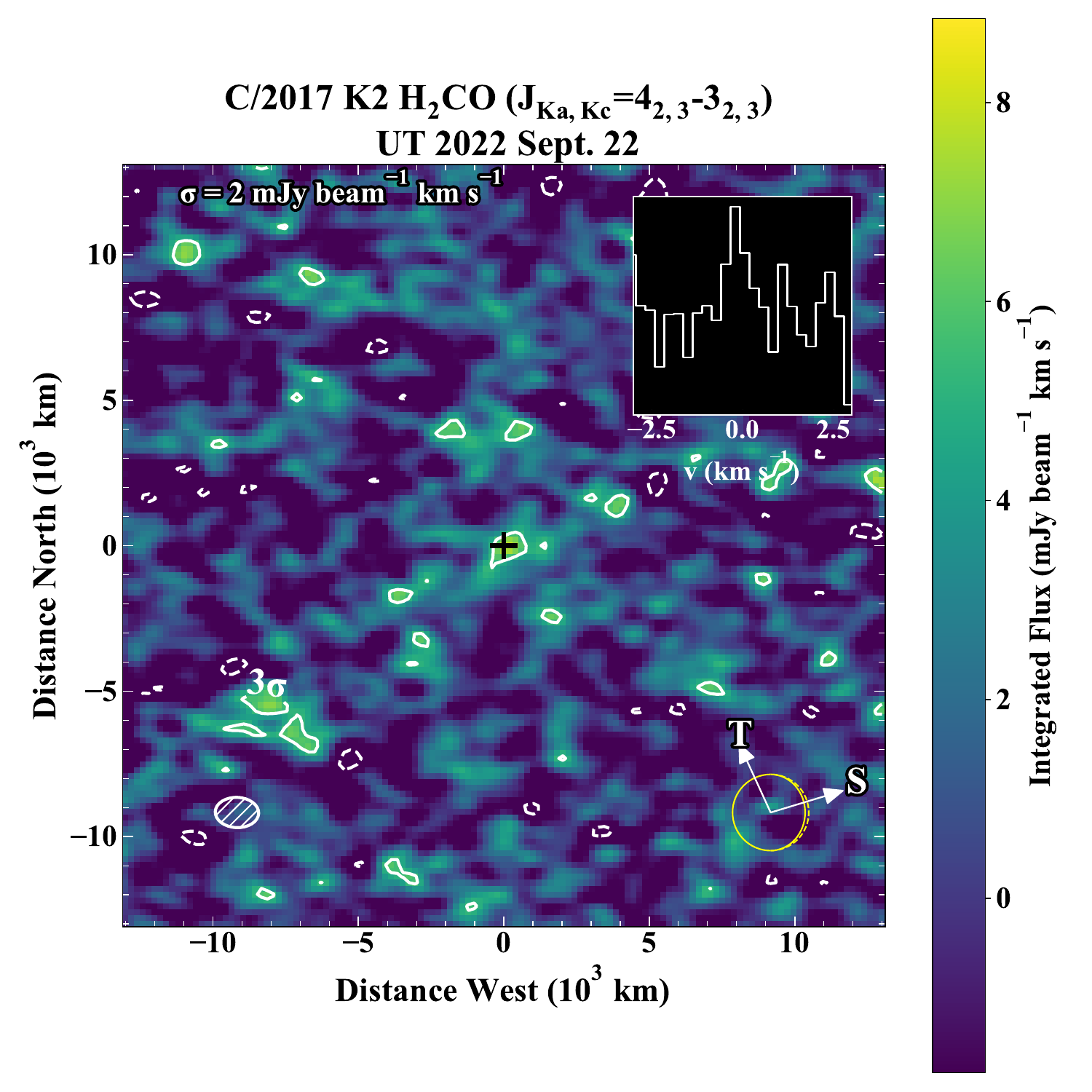}{0.3\textwidth}{(B)}
          \fig{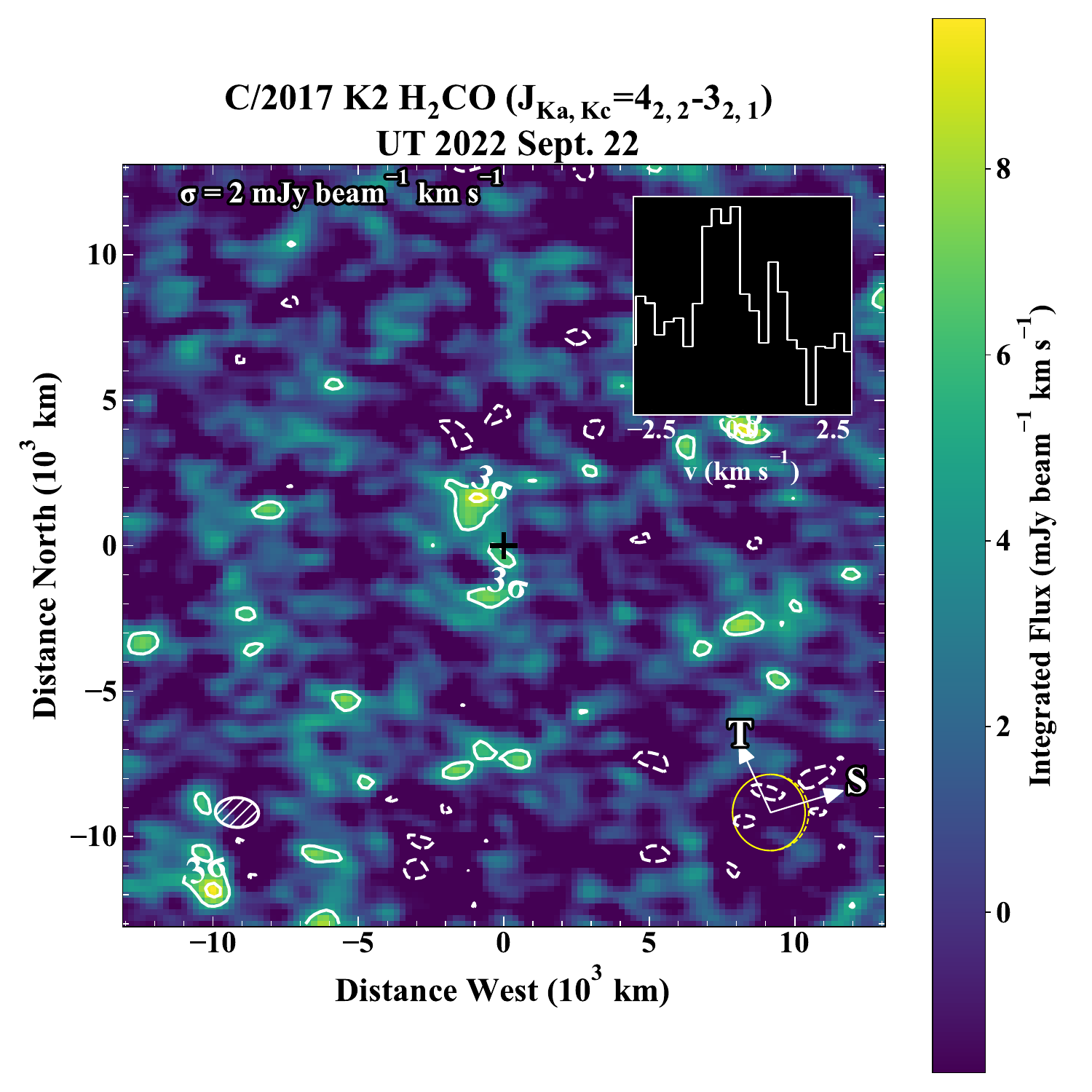}{0.3\textwidth}{(B)}
}
\gridline{\fig{K2.H2CO-351GHz.pdf}{0.3\textwidth}{(C)}
          \fig{K2.H2CO-362GHz.pdf}{0.3\textwidth}{(D)}
}
\caption{\textbf{(A)--(D).} Spectrally integrated flux maps for all detected H$_2$CO transitions in K2, with traces and labels as in Figure~\ref{fig:hcn-maps}. Contours are in 5$\sigma$ increments for H$_2$CO ($J_{Ka,Kc}=5_{1,5}-4_{1,4}$). Contours are 3$\sigma$ and $5\sigma$ for H$_2$CO ($J_{Ka,Kc}=5_{0,5}-4_{0,4}$), ($J_{Ka,Kc}=4_{0,4}-3_{0,3}$), ($J_{Ka,Kc}=4_{2,2}-3_{2,1}$), and ($J_{Ka,Kc}=4_{2,3}-3_{2,3}$). 
\label{fig:allH2CO}}
\end{figure*}

Here we provide integrated flux maps for all detected H$_2$CO transitions in comet C/2017 K2 (Figure~\ref{fig:allH2CO}).

\section{Fourier Domain Molecular Spectra}\label{sec:spectra}

\begin{figure}
\includegraphics[width=\textwidth]{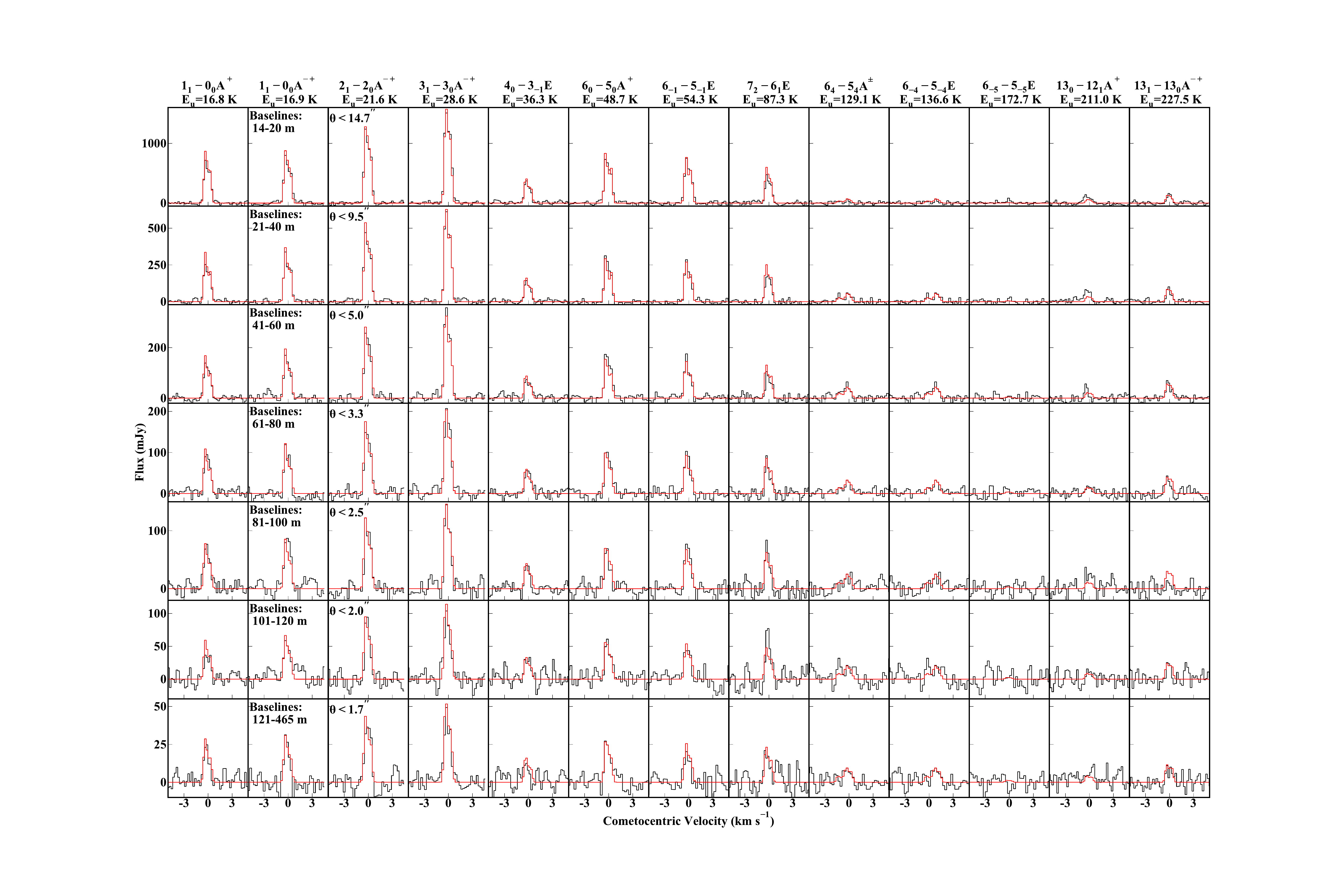}
\caption{Observed spectra for all detected CH$_3$OH transitions in C/2017 K2 with the best-fit model overlaid in red. Each spectrum is averaged over a range of $uv$-distances, provided in the upper left corner, corresponding to a range of angular scales in the coma.
\label{fig:ch3ohSpec}}
\end{figure}

\begin{figure}
\plotone{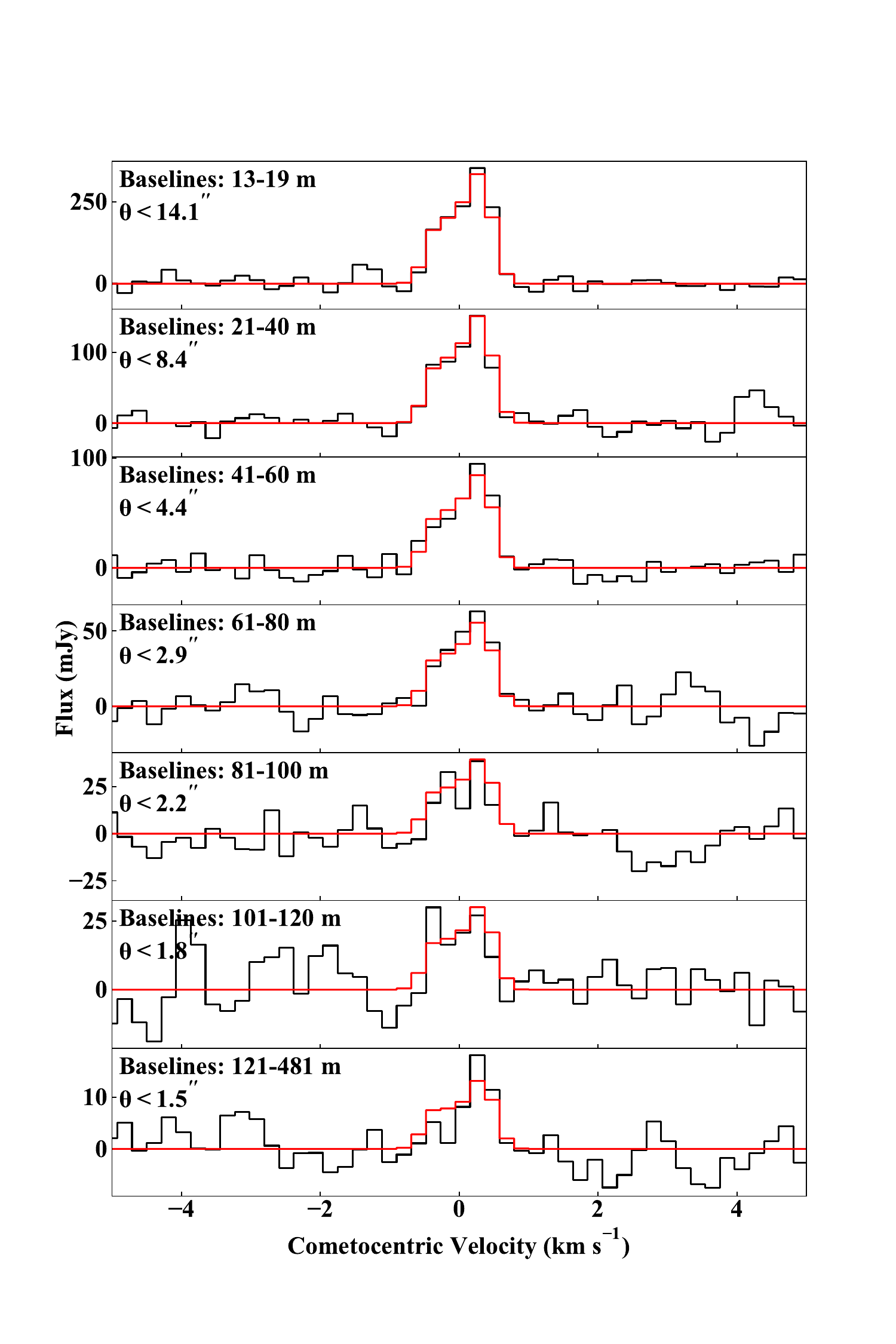}
\caption{Observed spectra for CO (\Ju{}=3--2) transition in C/2017 K2, with traces and labels as in Figure~\ref{fig:ch3ohSpec}.
\label{fig:coSpec}}
\end{figure}

\begin{figure}
\plotone{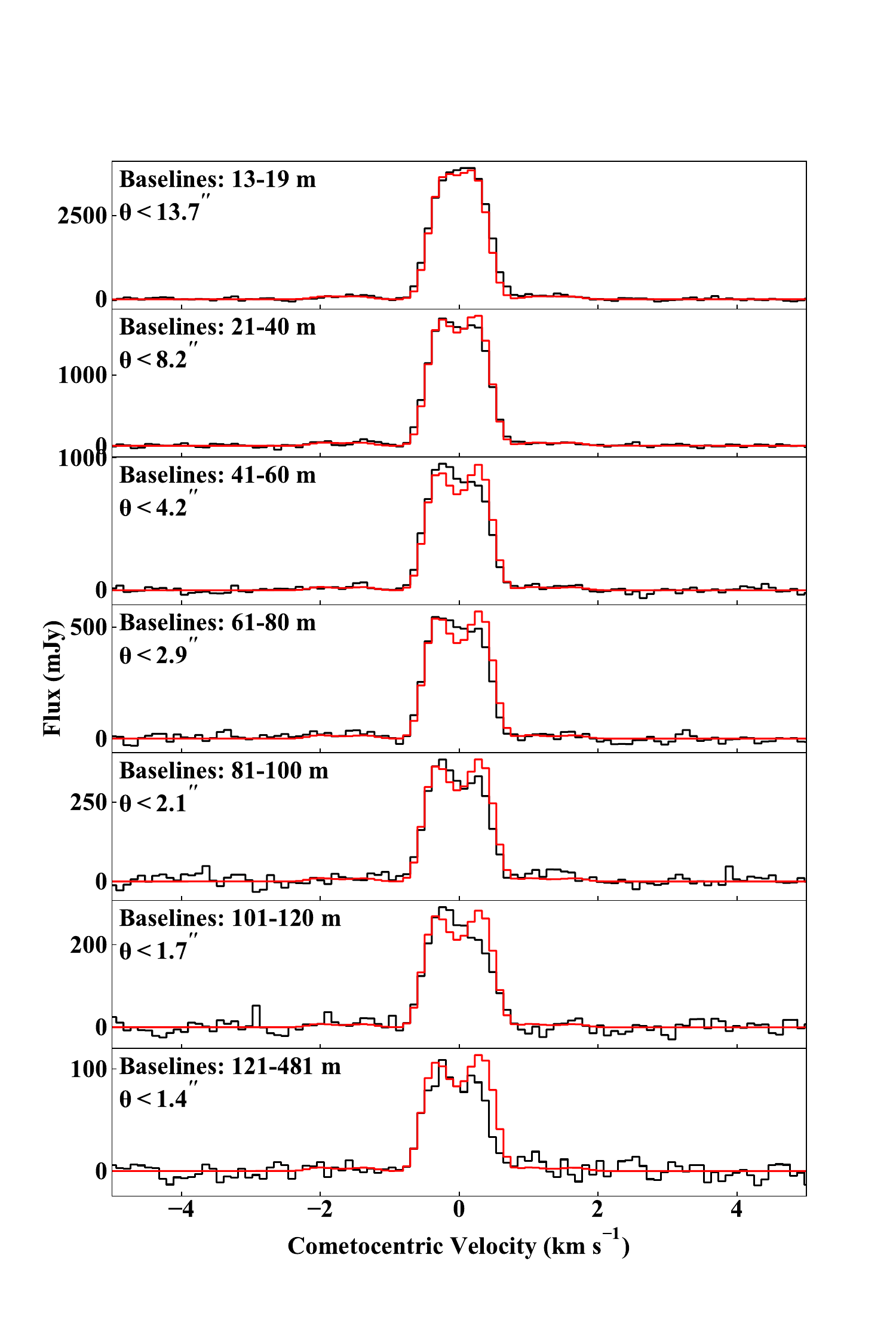}
\caption{Observed spectra for HCN (\Ju{}=4--3) in C/2017 K2, with traces and labels as in Figure~\ref{fig:ch3ohSpec}.
\label{fig:hcnSpec}}
\end{figure}

\begin{figure}
\plotone{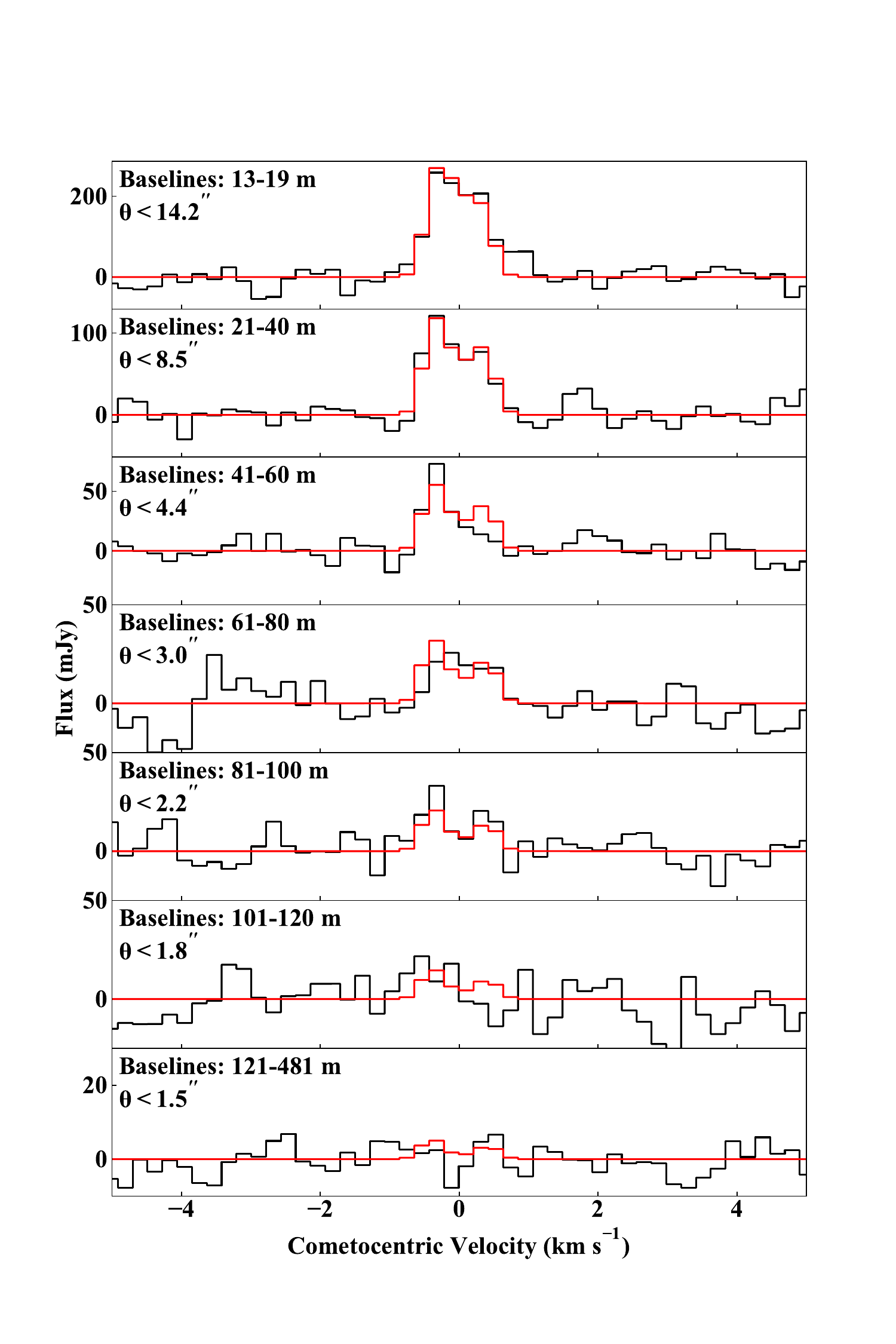}
\caption{Observed spectra for CS (\Ju{}=7--6) in C/2017 K2, with traces and labels as in Figure~\ref{fig:ch3ohSpec}.
\label{fig:csSpec}}
\end{figure}

\begin{figure}
\plotone{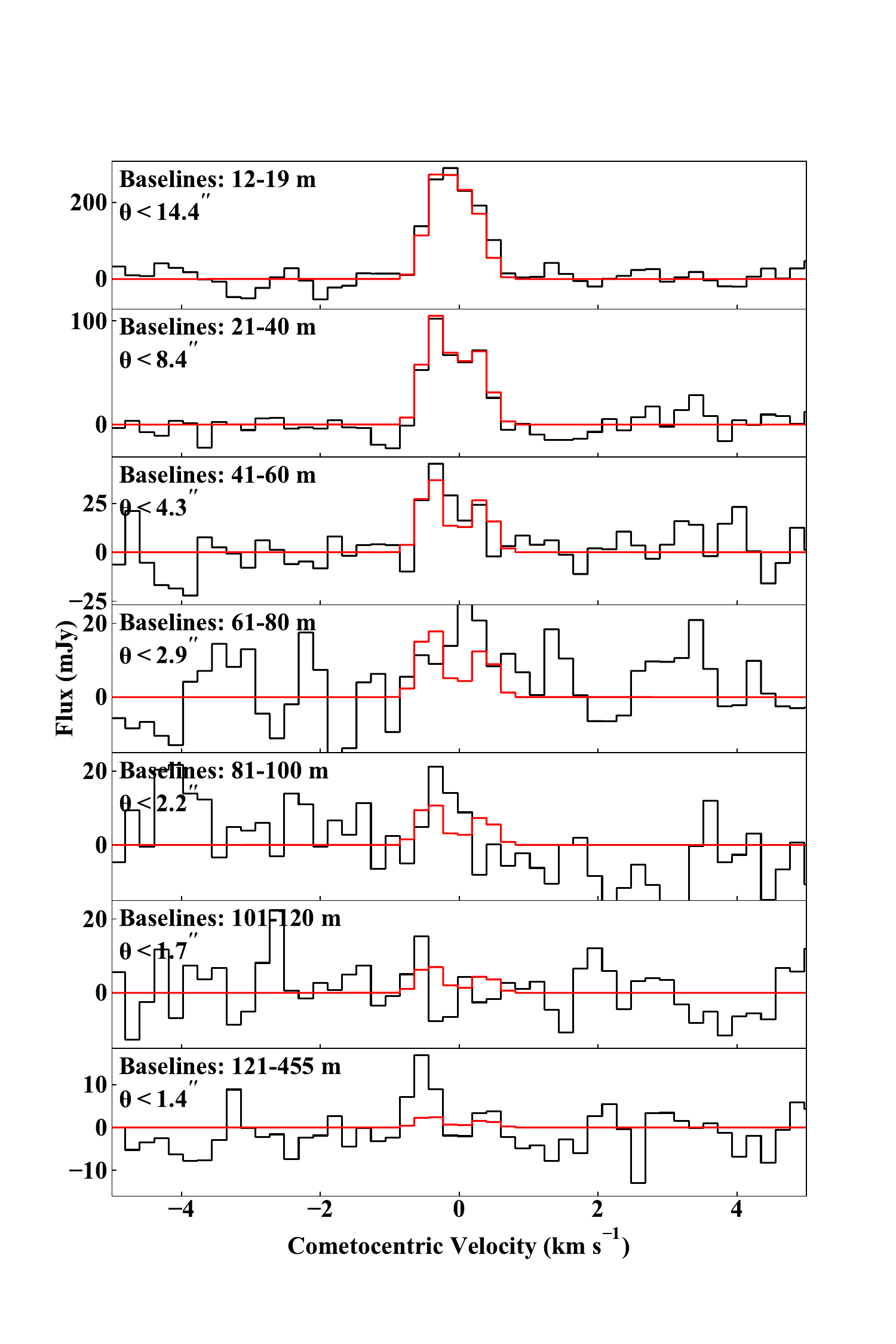}
\caption{Observed spectra for H$_2$CO ($J_{Ka,Kc}$=$5_{1,5}$-$4_{1,4}$) in C/2017 K2, with traces and labels as in Figure~\ref{fig:ch3ohSpec}.
\label{fig:h2coSpec}}
\end{figure}

Here we provide Fourier domain spectra for all detected species, providing a model-data comparison as a function of baseline and thus the variation in spectral intensity as a function of angular scale in the coma (Figures~\ref{fig:ch3ohSpec}-\ref{fig:h2coSpec}). 

\bibliography{K2}{}
\bibliographystyle{aasjournalv7}



\end{document}